\documentclass[sigconf,nonacm]{acmart}

\usepackage{xspace}
\usepackage{multirow}
\usepackage{multicol}
\usepackage{booktabs}

\usepackage{float}
\usepackage{listings}
\usepackage{xcolor}
\usepackage{graphicx}
\usepackage{amsmath}
\usepackage{subcaption}
\usepackage{wrapfig}

\lstdefinestyle{promptstyle}{
  basicstyle=\ttfamily\small,
  frame=single,
  breaklines=true,
  columns=fullflexible
}

\title{Forecasting Cybersecurity Incidents Using\\ Geopolitical Data and Large Language Models}

\newcommand{\parheading}[1]{\medskip\noindent\textbf{\textit{#1}}\hspace{2pt}}

\newcommand{\llm}[0]{LLM\xspace}
\newcommand{\csidb}[0]{CSIDB\xspace}
\newcommand{\cissm}[0]{CISSM\xspace}

\newcommand{\org}[0]{\ensuremath{\mathit{org}}\xspace}

\newcommand{\secref}[1]{\S\ref{#1}\xspace}

\newcommand{\figref}[1]{Fig.~\ref{#1}\xspace}
\newcommand{\tabref}[1]{Table~\ref{#1}\xspace}

\author{Mark Fesenko}
\authornote{Both authors contributed equally.}
\affiliation{\institution{Tel Aviv University}\country{Israel}}

\author{Abdullah Garra}
\authornotemark[1]
\authornote{Work done while at Tel Aviv University.}
\affiliation{\institution{University of Massachusetts Amherst}\country{USA}}

\author{Yaniv Harel}
\affiliation{\institution{Tel Aviv University}\country{Israel}}

\author{Mahmood Sharif}
\affiliation{\institution{Tel Aviv University}\country{Israel}}

\AddToHook{shipout/foreground}{%
  \ifnum\value{page}=1
    \put(54,-32){%
      \parbox{\dimexpr\paperwidth-1.5in\relax}{%
        \centering\textsc{\large Preprint}\\[2pt]\rule{\linewidth}{0.6pt}}}%
  \fi}

\begin{document}

\begin{abstract}
    Predicting security incidents is a profound task critical for informing proactive defensive measures and cyber-insurance policies. Prior work tackling this problem mainly utilized structured, manually defined features based on network measurements (e.g., protocol misconfigurations). Still, despite leading to promising performance, the network-based features may fail to capture aspects related to adversaries' motives.

To fill this gap, our work leverages geopolitical data mined from public sources--which may help capture attacker motives--to forecast security incidents. Specifically, our approach relies on news articles and transcribed podcasts that are fed to large language models to automatically produce rich representations. The representations are then fed to a classifier trained to forecast future incidents based on historical ones.

Our evaluation with a large incidents dataset ($>$15,700 records) demonstrates substantial accuracy (71.4\% ROC AUC) with geopolitical data alone. Notably, combining geopolitical data and network measurements outperforms the state-of-the-art technique based on network features alone (81.3\% vs.\ 77.5\% ROC AUC). Our analysis also helps shed light on when geopolitical data is most helpful and the data sources that are most useful for accurate forecasting.

\end{abstract}


\maketitle

\section{Introduction}

The increasing frequency and impact of cybersecurity incidents have
motivated growing interest in forecasting potential incidents before
they occur.
Predictive cyber-risk modeling aims to identify organizations that are
likely to suffer an incident within the near future~\cite{liu2015cloudy,Soska2014CompromisePredict},
allowing for timely mitigation and resource allocation.
Among others, such forecasts can help inform proactive defenses, for instance, by enabling
experts to prioritize vulnerability patching--a core tenet of risk-based vulnerability management~\cite{sabottke2015vulnerability}.
Beyond operational security, cybersecurity incident forecasting can also play a
critical role in guiding cyber-insurance policies and risk assessment~\cite{romanosky2016examining}.

Traditional methods for forecasting cybersecurity incidents primarily
rely on structured organizational and network-based features, such as
untrusted certificates, DNS port randomization, exposed services, or
vulnerability metrics~\cite{liu2015cloudy,Sharif18Prediction,Soska2014CompromisePredict}.
A notable example is the work of Liu et
al.~\cite{liu2015cloudy}, which demonstrated that combining public
vulnerability and (mis)configuration data can yield meaningful predictive
signals for future attacks.

Still, although such structured data provide an important foundation, they
capture only part of the picture. In practice, much of the contextual
information surrounding an organization's cyber posture--such as news
coverage, partnerships, acquisitions, public exposure, or even
mentions of past security issues--appears in unstructured text
sources. Moreover, while public network indicators may hint at the
technical feasibility of attacks, they may not be able to capture
attacker motives, as, for example, geopolitical data and news articles
may capture~\cite{Hunter21AttackFactors}.
Here, advances in large language models (\llm{}s)~\cite{brown2020language,Vaswani17Attention} provide new
opportunities to extract latent patterns from such textual data at
scale, potentially revealing precursors to breaches that are not
reflected in technical telemetry alone.

Going beyond traditional approaches,
we propose a framework that  leverages geopolitical data
and can further integrate unstructured text from articles with
structured network features to predict future cybersecurity
incidents.
Using an \llm{}-based text encoder (\secref{sec:forecasting_system}), we transform news articles (e.g.,
ones mentioning target organizations) into dense semantic
representations (i.e., vectors) that represent the surrounding
context.
These representations can then be used alone or combined with network
features in a joint classification models for forecasting incidents.
In essence, similarly to prior work that argues that lacking organizational security hygiene is indicative of more severe security vulnerabilities likely to be exploited by adversaries~\cite{liu2015cloudy}, our work suggests that geopolitical news encode information (e.g., geopolitical tensions or public outrage) that are correlated with security incidents.
Our results indicate that our proposed approach based on \llm{}s using
geopolitical data leads to forecasts with competitive accuracy in
comparison to previous methods based on network indicators.
Moreover, the incorporation (i.e., fusion) of textual representations
and network features (\secref{sec:forecasting_system:fusion}) significantly
improves prediction performance
compared to network-only features,
highlighting the complementary value of geopolitical data and
technical indicators.

\medskip{}\noindent
In a nutshell, our contributions are fourfold:
\begin{itemize}
    \item We enrich a popular cybersecurity-incident dataset covering $>$15,700
    incidents from the past 12 years~\cite{harry2018classifying} (\secref{sec:data:incident}), improving the
    accuracy of the reported incident dates, thus ensuring reliable model
    training and assessment.

    \item We collect and pre-process large-scale geopolitical dataset,
    consisting of news articles and podcasts, spanning over a decade, and reaching
    up to 2.5 million monthly records (\secref{sec:data:geopol}). Moreover, we
    gather network features to
    reproduce seminal work by Liu et al.\ on predicting data breaches~\cite{liu2015cloudy}
    and augment the input of our forecasting system (\secref{sec:data:network}).

    \item We design an \llm{}-based system that ingests relevant geopolitical
    data to produce rich representations for forecasting security incidents (\secref{sec:forecasting_system}).
    Crucially, our system is extensible, enabling fusion between geopolitical data
    and network features to improve forecasts' accuracy.

    \item We extensively evaluate our approach (\secref{sec:evaluation}),
    demonstrating the viability of  leveraging geopolitical data for forecasting
    security incidents, and advancing the state-of-the-art by fusing geopolitical
    and network data for more accurate predictions. Furthermore, we uncover the
    settings where geopolitical data is most useful (e.g., predicting incidents
    motivated by protests), and characterize the types of inputs that enable
    predictions (e.g., popular domains are a useful data source for predictions).
\end{itemize}

The remainder of this paper is organized as follows:
We next review related work on forecasting events in cybersecurity
and other domains (\secref{sec:relwork}) before laying out our
threat model (\secref{sec:threat_model}).
Then, we describe our data-collection apparatus (\secref{sec:data}),
our forecasting system's design (\secref{sec:forecasting_system}),
and the limitations of our work (\secref{sec:limits}).
Finally, we wrap the paper with the evaluation results
(\secref{sec:evaluation}) and a conclusion (\secref{sec:conclusion}).

\section{Related Work}
\label{sec:relwork}

In this section we present prior work on forecasting future
events in computer security (\secref{sec:litrev:sec}) and other domains (\secref{sec:litrev:nonsec}).

\subsection{Forecasting Security Incidents}
\label{sec:litrev:sec}

Prior predictive security analytics work has primarily relied on
system and behavioral measurements (e.g.,~\cite{liu2015cloudy,Sarabi2016BreachPredict,Sharif18Prediction,Soska2014CompromisePredict}) and self-reported data
(e.g.,~\cite{Sharif18Prediction}) to collect observations and enable
predictions. For instance, Soska and Christin collected information
about content-management system versions and installed plugins to
forecast whether web servers will be compromised within a year from data
collection~\cite{Soska2014CompromisePredict}. As another example,
Sharif et al.\ gathered Internet browsing data (e.g., website
categories and amount of bytes downloaded or uploaded), both
historical and within browsing sessions, as well as users' responses
to standard questionnaires to predict impending exposure to malicious
content on the fly, as users are browsing~\cite{Sharif18Prediction}.

Most related to our work is the work of Liu et
al.~\cite{liu2015cloudy}.
Their work aimed to forecast organizations' security breaches by leveraging
externally observable properties of the organizations' networks.
The researchers engineered and compiled different features falling into two main categories: mismanagement symptoms (such as misconfigured DNS or BGP records) and
malicious activity time series (including spam, phishing, and scanning activity).
Subsequently, using a machine-learning model, they demonstrated high prediction accuracy.

The features used for prediction in prior work
(e.g.,~\cite{liu2015cloudy,Sarabi2016BreachPredict,Sharif18Prediction,Soska2014CompromisePredict}) were mostly hand
crafted by the researchers, requiring a laborious feature engineering
and domain knowledge.
Moreover, because humans may fail to discover specific correlations in
the data, they may exclude important features conducive for accurate
predictions. Not to mention that the manual work required for
developing features may preclude timely updates to the system as new
attacks emerge.

Differently from prior work, our work strives to leverage
geopolitical news to forecast security incidents.
Publicly available geopolitical data complement observations
used in older systems in that they encode information that describe
attacker motives and are correlated with security incidents.
For instance, national conflicts extensively discussed in the news are often followed by cyber attacks conducted by nation states or associated parties (e.g.,~\cite{RU-UA-powergrid}).
Since motives play a central role in
attacks~\cite{Hunter21AttackFactors}, we expect
that our work would inform analytics systems producing high-quality
forecasts.
As geopolitical information is usually
available in the form of natural language, we build upon adequate
techniques to extract features necessary for prediction. Namely,
toward addressing the shortcomings of prior work that relied on
manual feature engineering, we incorporate automatic
feature-extraction methods through \llm{}s~\cite{Vaswani17Attention}.

Beyond predictive analytics, the use of \llm{}s in cybersecurity has
expanded rapidly. The systematic review by Zhang et
al.~\cite{zhang2025llms} catalogs hundreds of works applying \llm{}s
to vulnerability detection, malware analysis, threat intelligence,
and incident response. Jones et al.~\cite{jones2025analysing} provide
a complementary view focused on incident management, evaluating how
different \llm{} families support each stage of the response
lifecycle. Both reviews, however, treat \llm{}s primarily as reactive
tools applied after an incident has been observed or a vulnerability
disclosed. To the best of our knowledge, the use of \llm{}s in predictive
instruments that anticipate attacks before they occur was not previously studied.

\subsection{Forecasting Events Unrelated to Security}
\label{sec:litrev:nonsec}

A significant body of work
studies the
problem of forecasting events in domains other than computer
security, ranging from science~\cite{Ji24Citation,Muralidhar20Physics} to
health~\cite{Adhikari19Epidemic,Chakraborty18Flu,Rekatsinas17RareDisease,Rodriguez21Flu,Roy2021Covid}, from
different world events (e.g., civil
unrest)~\cite{Hossain22WorldEvents,Korkmaz16Unrest,Yan23Autocast++,Zhao21WorldEvents,Zou22WorldEvents} to social
networks~\cite{Islam19OSN,Zhao17OSNs}, and
covering other domains as well~\cite{Akhter18Homicides,Chen20Traffic,Jin17Stocks,Keneshloo16NewsPop,Khadivi16Tour,Khandpur17Airport,Muralidhar19CPS}.
Past work
in domains other than computer security
has primarily considered specific time scales for predictions (e.g.,
months~\cite{Adhikari19Epidemic}) while offering little flexibility for changing these scales;
mainly relied on manually crafted time-series and historical features (e.g.,~\cite{Rodriguez21Flu});
and mostly used dated (pre-transformer architecture) models (e.g.,~\cite{Muralidhar20Physics,Muralidhar19CPS}).
In contrast, our work, besides tackling predictions in computer security,
leverages automatically extracted features from text;
employs state-of-the-art transformer-based models;
and enables forecasts at varying times-scales.

The recent efforts of Turtel et al.~\cite{turtel2025outcome} and
Wang et al.~\cite{wang2024news} are notable exceptions to these trends.
The former shows how to fine-tune \llm{}s to predict betting results.
The latter uses \llm{}-based agents to filter and
analyze news articles, integrating the resulting event
representations with numerical time series to produce forecasts
across several domains, such as predicting stocks and cryptocurrency
prices. They demonstrate that text-derived event signals improve accuracy
over time-series-only baselines. Our work adapts this high-level recipe
to organizational cyber risk: we pair dense textual representations of
geopolitical events and organization-related news with structured Internet-measurement telemetry, and evaluate under time-realistic conditions across multiple forecast horizons, a combination that appears underexplored in prior non-security forecasting work.

\section{Threat Model and Problem Formulation}
\label{sec:threat_model}
Our goal is to assess the risk that an organization will experience a cybersecurity incident within a given, pre-defined time window into the future. To this end, we parameterize the notion of cybersecurity incidents.
\begin{definition}
  $\tau$-cyber incident: Consider an organization $\org$ at time
  $t$. The organization is deemed to suffer from a $\tau$-cyber
  incident if such incident is reported within time $\hat{t}$ such
  that $t\le\hat{t}\le{}t+\tau$.
\end{definition}
With this definition, our formulation follows prior work on organizational data-breach forecasting~\cite{liu2015cloudy} in using binary outcome labels over a fixed forecasting horizon, while modeling incident risk as a continuous score. Formally, we define the cybersecurity-incident forecasting problem as follows:

\begin{definition}
  Cybersecurity-incident forecasting:
  For a given organization $\org$ at time $t$, the forecasting
  function $F_\tau$ receives observations $o_{\org}$ and outputs a risk score $F_\tau(o_{\org})\in [0,1]$ assessing the risk of a $\tau$-cyber incident occurring.
\end{definition}

In practice, the output of $F_\tau(o_{\org})$ may be binarized according to
some threshold depending on a desired operating point, thereby yielding a forecasting system aligned with the operator's risk tolerance and tradeoffs. For a given threshold, we say that the binarized output of the system is accurate if it
outputs 1 (resp.\ 0) and an actual incident occurs (resp.\ does not
occur) within $[t, t+\tau]$.
In our research, we primarily consider a prediction window of $\tau$=30 days, but we also evaluate
$\tau$=1 year.
These durations are consistent with prior work~\cite{liu2015cloudy},
and provide a window of opportunity for applying proactive defensive measures.

\parheading{Cybersecurity Incidents.}
We aim to predict major cybersecurity incidents, akin to those
reported in public media and recorded in
cybersecurity-incident databases~\cite{harry2018classifying,csidb}.
Such cyber incidents include, but are not limited to, denial of
service attacks, data exfiltration, ransomware, or even physical
attacks (e.g., manipulation of physical controllers or power
grids)~\cite{harry2018classifying,csidb}.
Note that, in certain jurisdictions and under specific
conditions, regulations may
obligate organizations to report security incidents that they suffer
from.
For instance, reporting is typically required for incidents that involve a
high risk to the rights of individuals (e.g., European law requires notification
of personal data loss within 72 hours of discovery~\cite{Art33GDPR}),
or those that have a
material impact on public safety or financial markets (e.g., US regulations require
reporting substantial cyber incidents within 72 hours and any ransomware payment
within 24 hours of payment~\cite{10.1108/OXAN-DB282275}).
Such regulations help make our data more complete~\cite{harry2018classifying,csidb},
thus enabling more
reliable evaluation of forecasts' accuracy.

\parheading{Observations (I.e., $F_\tau$'s Inputs, $o_{\org}$).}
The predictive model, $F_\tau$, operates solely on publicly available
information, primarily derived from public media. It does not assume
access to internal telemetry or attacker-specific data.
For each organization, the forecasting model $F_\tau$ receives contextual
inputs composed of several information sources to enable predictions.
As explained in \secref{sec:litrev:sec}, prior work leveraged
technical indicators, such as DNS or BGP misconfigurations and
appearance of organizational devices on blacklists, to forecast
incidents~\cite{liu2015cloudy}.
In contrast, we focus on organization-related, geopolitical news collected from the
written press and transcribed podcasts segments, forming geopolitical
indicators at the level of the organization, its sector, or its
country, to inform predictions.
For predictions at time $t$, the geopolitical observations are collected from a historical time
window $\in[t-h, t]$, and are later fed into an \llm{} (as part of
$F_\tau$) to extract numeric features that are then mapped to predictions.
For best performance, $h$ can be tuned so that it is large enough to
capture relevant information, but small enough so that the information
is useful for predictions in the given time horizon $[t, t+\tau]$.

We emphasize that $F_\tau$ is trained to exploit statistical correlations between observed inputs and subsequent incidents, rather than to recover causal relationships between them. This framing is shared with previous work using network-indicator-based forecasters~\cite{liu2015cloudy}: the presence of misconfigurations in an organization's network does not itself constitute a vulnerability, but indicates an absence of the policies needed to detect and remediate such failures, under which breaches are more likely. Analogously, a surge in geopolitical friction surrounding an organization's country or sector does not directly cause an attack against that organization, but may indicate conditions (such as heightened adversary motivation, strained diplomatic channels, or retaliatory cycles) under which attacks become more likely.

\section{Data Collection and Processing}
\label{sec:data}

To train and evaluate forecasting models, we need both incident and
observation data. The incident data serve as the dependent variables
that the model seeks to predict, whereas the observation data serve as
the independent variables. In both cases, as previously explained
(\secref{sec:threat_model}), we rely on publicly available data.
Doing so helps in maximizing reproducibility and transparency, and in
obtaining broad, vendor-agnostic coverage without privileged access or
intrusive measurement. For
cybersecurity incidents (\secref{sec:data:incident}), we use public
repositories recording security incidents discussed in the press.
For observations, we primarily use geopolitical data from news articles and podcasts
(\secref{sec:data_pipeline}), and consider augmentations with
summary reports of global events (\secref{sec:gdelt}) and
network measurements (\secref{sec:network_indicators}).

\subsection{Cybersecurity Incident Data}
\label{sec:data:incident}

We use the Center for International and Security Studies at Maryland
dataset (\cissm)~\cite{harry2018classifying} as the primary source of
incident data. \cissm{} offers longitudinal, organization-level
records of publicly reported cybersecurity incidents (2014 onward)
curated under a consistent taxonomy and with public provenance (i.e.,
with references to public articles about incidents), making it
well-suited for training and evaluating organization-level forecasting
models. To date, \cissm{} contains $>$15,700 records about past
incidents
.
Each record includes the organization name, event type, and a reported
incident date, and, when available, information about the actor's type
and  motivation, and the organization's sector.

A practical caveat of \cissm{}, however, is that some records inherit the
publication dates of news articles discussing incidents in lieu of the (earlier)
incident dates stated in the articles' body.
To address this limitation, we used the Cyber Security Incident
Database (\csidb)~\cite{csidb} as a corrective layer over \cissm, as
we found it mostly agrees with \cissm{} on incident dates, but is more
accurate when it reports different (usually earlier) dates.
Operationally, we matched \cissm incidents and with corresponding
incidents in \csidb via the cited source article. When a match was
found, we replaced the \cissm timestamp with the incident date
reported by \csidb.
Among the 4,700 incidents whose timestamps were corrected using CSIDB (29.8\% of all incidents), the median correction was 25 days, further motivating the choice of $\ge$30-day prediction window in our experiments (\secref{sec:evaluation}). A manual audit  of 48
records and their corresponding articles
, showed that this
process increased the accuracy of the incident date field from 77.1\% to
91.7\%.

\subsection{Geopolitical Data Collection Pipeline}
\label{sec:data:geopol}
\label{sec:data_pipeline}
High-quality and high-coverage news data is essential for the
forecasting model we use. Thus, we prioritized building an accurate
and efficient data-ingestion pipeline based on large-scale data from
reliable sources.

\parheading{News Articles}
We collect our raw data from the CC-NEWS dataset, a continuously
updated archive of global news extracted from Common
Crawl~\cite{ccnews}. Each monthly snapshot comprises of articles' text
(specifically, HTML pages) scraped from news sources across the
web. These snapshots span multiple years, from January 2016 onward,
and offer a rich, multilingual resource for temporal and geopolitical analysis.

Monthly volumes vary over time, peaking at approximately 2.5M monthly articles.

We observed that, in every year, the maximal percentage of any single source is just below
4\% of that year's articles, indicating that the corpus remains diverse across years.

\parheading{Podcasts}
To broaden the scope of organization and cyber-related media coverage, we collected
a dataset of podcasts designed to capture both technical nuances and broader
contextual signals.
We selected 16 distinct sources categorized into two groups:
(1) \emph{Domain-specific cybersecurity outlets} (e.g., Darknet Diaries
and The CyberWire) to provide granular details on vulnerabilities, incident
forensics, and threat actor TTPs; and
(2) \emph{General news and geopolitical sources} (e.g., The Wall Street Journal,
BBC, and NPR) to capture the economic instability and political friction
that often precipitate major cyberattack campaigns.

This variety ensures coverage of distinct layers of cyber and geopolitical events,
extending from broad economic and political discussions to the reporting of emerging
exploits and cyberattacks.

While we collected fewer podcast episodes than news articles, our podcast collection
remains substantial in absolute terms, spanning diverse sources over a significant
timeframe.

In total, we collected over 30,000 podcast episodes, amounting to more than 14,000 hours of audio, each with a varied number of episodes,
where the duration of each podcast varies from a just a few minutes up to nearly
200 minutes.
These episodes span multiple years, going back as far as 2016 and reaching recent days
in 2025.

\parheading{Data Ingestion and Pre-Processing}
\label{sec:data_ingestion}
To efficiently process and extract useful content from new articles'
HTML files, we utilized the popular DataTrove
framework~\cite{penedo2024datatrove}, which implements pipelines
for distributed processing of HTML files.

Specifically, we used DataTrove's boilerplate removal
capabilities (based on Trafilatura~\cite{barbaresi2021trafilatura}) to extract
news articles' text while removing irrelevant content from the raw HTML
files.

Moreover, using DataTrove's language detector, we retained only articles written
in English, as English is best supported by the state-of-the-art \llm{}s that
our forecasting system utilizes.

Last, for each article, we extracted the publication date via
DataTrove. When  unavailable, we defaulted to the CC-News's timestamp
otherwise. The resulting records were then inserted into a relational
database to facilitate faster querying and linkage with other datasets.

Furthermore, we transcribed each podcast episode to enable subsequent
analysis using \llm{}s. Specifically, we employed the open-source
Whisper-large model~\cite{radford2023robust} for transcription, finding it to be
competitive with commercial alternatives
. We cleaned and segmented transcripts into smaller chunks, with a maximum
length determined by an embedding model's context limit (see
below).
Each segment was then indexed individually to enable later retrieval.
We also added the transcripts to
the relational database, complementing it with data and discussions
that may be absent from traditional reporting (i.e., news articles).

To quantify transcription quality, we benchmarked Whisper-base and
Whisper-large against the commercial Deepgram~\cite{deepgram} API on three podcast
sources with human-reviewed reference transcripts (including \emph{The
New York Times} and \emph{99\% Invisible}).
Across match error rate
(MER$\downarrow$), word information lost (WIL$\downarrow$), word information preserved (WIP$\uparrow$),
and word error rate (WER$\uparrow$), Whisper-large achieved the best results
(5.42\% MER, 7.15\% WIL, 92.85\% WIP, 5.61\% WER), outperforming both
Deepgram (7.21\%, 9.61\%, 90.39\%, 7.39\%) and Whisper-Base
(8.24\%, 11.70\%, 88.30\%, 8.45\%).

Finally, all articles collected from CC-NEWS and the podcast
transcriptions were encoded using a pretrained text embedding model to
enable semantic search. Specifically, the embeddings produced by the model
can be used to retrieve text relevant to specific search queries,
subsequently augmenting the input we feed into our \llm{}
(\secref{sec:forecasting_system}). We selected \texttt{Arctic-embed-m-v2.0}
~\cite{yu2024arctic} as our embedding model because
it supports long context windows (up to 8,192 tokens)
, is relatively lightweight, and achieves strong retrieval results~\cite{muennighoff2023mteb}.

To support temporal retrieval, we stored the embeddings in a FAISS vector
database~\cite{douze2025faiss} partitioned by month.

Our results (\secref{sec:evaluation:results}) could, in principle, be sensitive to the choice of the retrieval model: different embedding models might surface different passages, leading to different \llm{} representations. We argue that this sensitivity is limited in practice for two reasons. First, \texttt{Arctic-embed-m-v2.0} is a top-performing retriever on the MTEB benchmark~\cite{muennighoff2023mteb} across both retrieval and reranking tasks, and has been shown to generalize well to long inputs~\cite{yu2024arctic}. Second, recent work on the Platonic Representation~\cite{huh2024platonic} argues that sufficiently capable embedding models converge to similar representations of the same underlying content, meaning that swapping one high-quality retriever for another empirically tends to surface largely overlapping passages~\cite{caspari2024beyond}. Accordingly, we expect our findings to remain stable across alternative high-quality dense retrievers.

\subsection{GDELT Contextual Data}
\label{sec:gdelt}
Besides raw geopolitical news data, we sought to incorporate
features to improve forecasts' accuracy (\secref{sec:forecasting_system:fusion}).
As one source of features, we used the Global Database of Events, Language,
and Tone (GDELT)~\cite{leetaru2013gdelt}--a large-scale open-source
dataset that tracks worldwide news media, encoding information about
events, entities, and tone from millions of sources in near real time.
In lieu of using the raw data, we queried the hosted version of GDELT to
extract specific aggregated indicators for each country and time period.
Specifically, we extracted features related to media attention and volume (total mentions
and total events), sentiment and stability (average tone and
Goldstein scale), and specific geopolitical event types (material conflict
events, protest events, and cooperation events).

\begin{figure}[h!]
  \centering
  \includegraphics[width=0.45\textwidth]{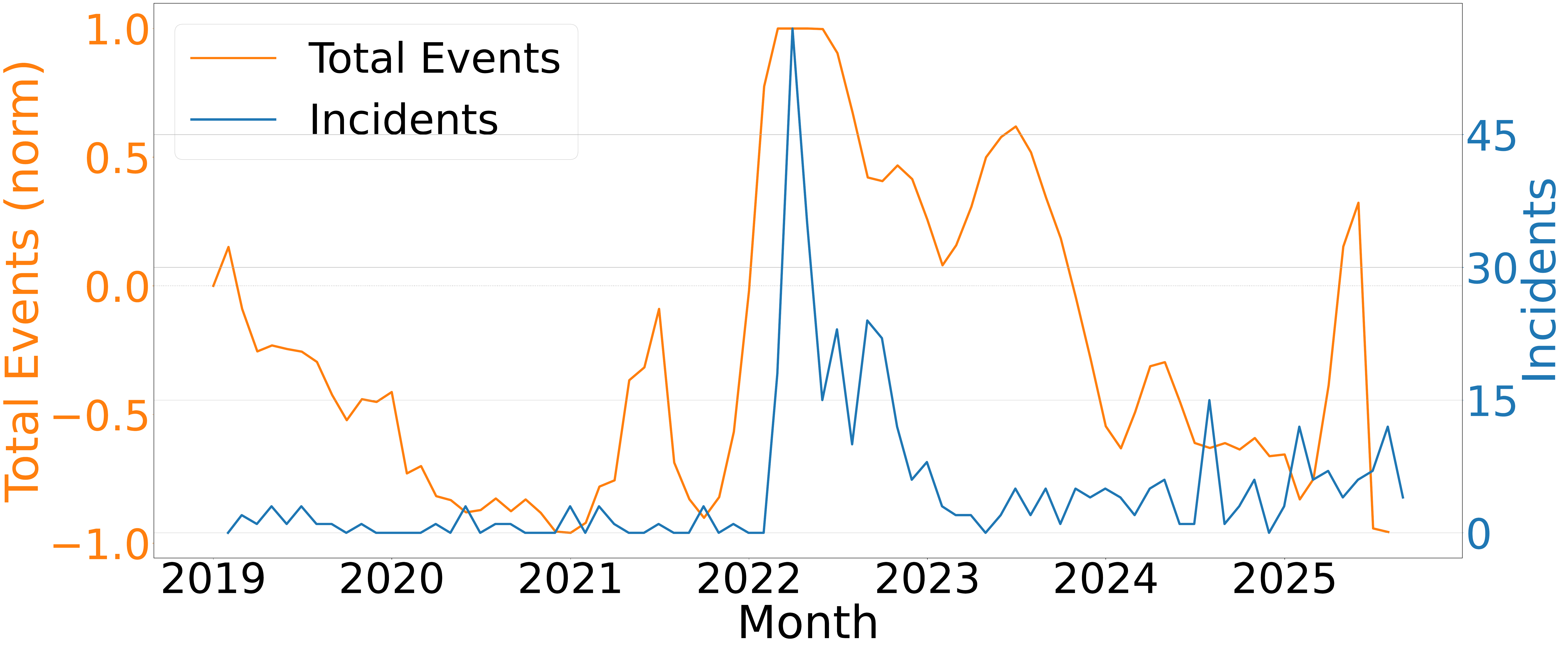}
  \includegraphics[width=0.45\textwidth]{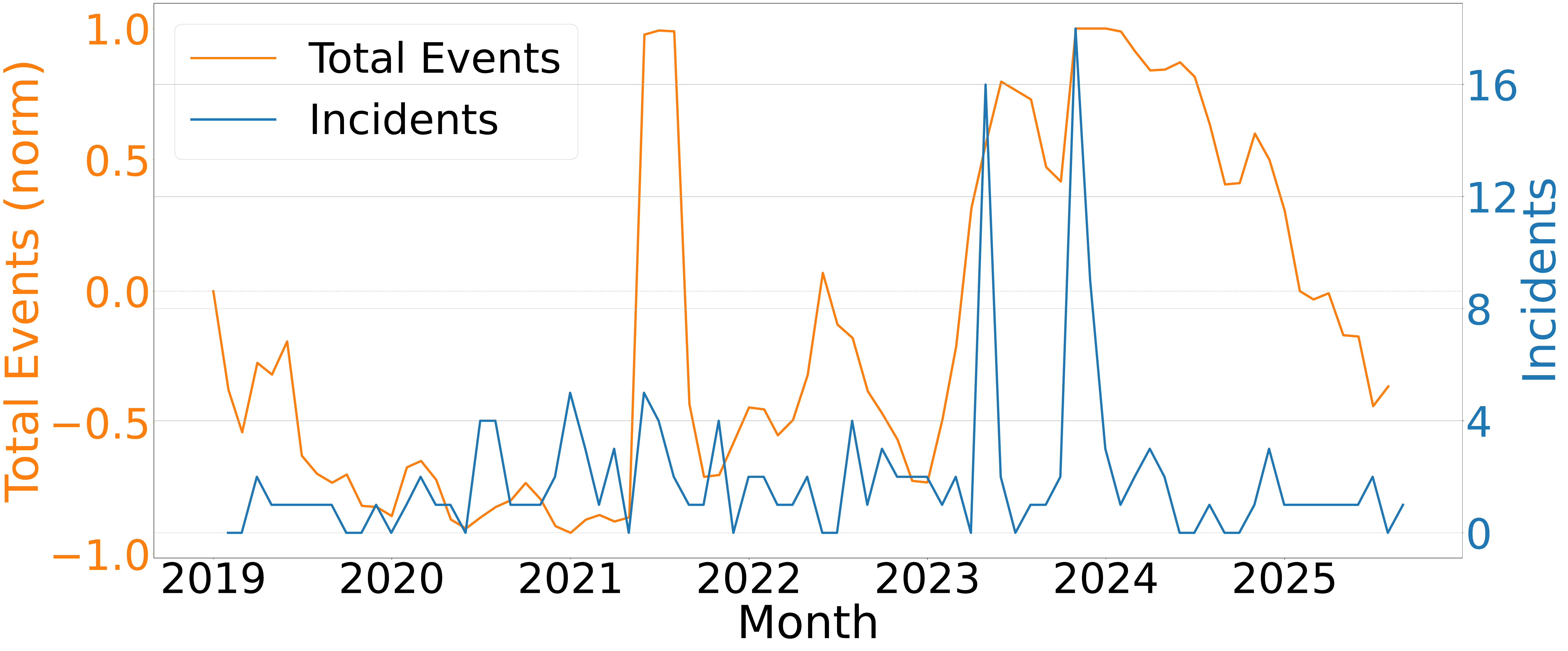}
  \caption{\label{fig:gdelt}The correlation between normalized conflict-event indicators derived from GDELT (orange curves) with the number of monthly security incidents (blue curves) in Russia (left) and Israel (right). The strong correlation between the indicators and incidents enable improved forecasts.}
\end{figure}

For each country and month, we transformed each raw indicator into a deviation-from-baseline representation. We first computed a $z$-score relative to the previous 30 months, excluding the current month. We chose a 30-month window as a balance between shorter windows, which are dominated by short-term noise, and longer windows, which dilute meaningful shifts in a country's geopolitical posture. We then applied a $tanh$ transformation to the $z$-scores,
bounding each feature to the $(-1, 1)$ range for later use as
numerical features, and limiting the influence of extreme outliers
that can arise from sparse months or reporting anomalies in GDELT. In the resulting representation, values near zero indicate typical activity, while positive and negative values indicate above- and below-baseline activity, respectively. As can be seen from \figref{fig:gdelt}, the normalized indicators derived from GDELT per are correlated with the number of monthly security incidents reported in certain countries.

To make this representation consumable by the \llm{}, we converted
each $z$-score into a short contextual phrase. We discretized each
variable into five intensity levels corresponding to
standard-deviation bands of the $z$-scores: \textit{unusually low}
($z \leq -1.0$), \textit{below normal} ($-1.0 < z \leq -0.33$),
\textit{typical} ($-0.33 \le z < 0.33$), \textit{above normal}
($0.33 \leq z < 1.0$), and \textit{unusually high} ($z \geq 1.0$).
We combined each intensity label with a coarse trend indicator to
produce phrases such as \textit{``Media attention toward the country
was unusually high and continued to rise in recent months.''} This
yields a compact, interpretable textual summary of a country's
recent geopolitical posture that can be appended to the prompt.

\subsection{Network Indicators}
\label{sec:data:network}
\label{sec:network_indicators}

We collected network mismanagement and malicious-activity data following Liu et al.~\cite{liu2015cloudy}, both to construct a baseline forecasting approach that reproduces their predictor as faithfully as possible in our setting—nearly a decade later and with more than 10$\times$ as many incident records-and to augment our geopolitical features with additional indicators for improved forecasting (\secref{sec:forecasting_system:fusion}).
Specifically, we collected two complementary data types that describe
organizations' externally exposed posture:
\emph{(1) mismanagement symptoms} (static configuration features) that
reflect longer-lived weaknesses in deployed infrastructure and policy,
and \emph{(2) daily malicious-activity signals} (time series).

\parheading{Mismanagement Symptoms}
We collected four static, mismanagement symptom features used by
Liu et al.~\cite{liu2015cloudy}.
These features capture information about:
\emph{Open Recursive Resolvers}, identified via a December 2024 \emph{Censys}~\cite{Censys}
DNS snapshot (finding 4.8 million open resolvers);
\emph{insufficient DNS source port randomization}, where we measured
the standard deviation of UDP source ports used by resolvers in
response to 10 queries, establishing a criterion to distinguish proper
from poor randomization;
\emph{untrusted HTTPS Certificates}, where we
scanned public HTTPS endpoints from the December 2024 \emph{Censys}
snapshot with \emph{ZGrab}~\cite{Censys,zgrab}, observing that 11.7
million IPs ($>$40\%) presented untrusted certificates;
\emph{BGP misconfigurations (short-lived routes)}, where we analyzed
BGP updates from 12 Route Views listeners over the December 2024 period
and tracked the lifetime of newly announced routes~\cite{routeviews,liu2015cloudy}.
Following prior work~\cite{mismanagement}, we treat routes that persist
for less than 24 hours as misconfigurations, identifying more than 1.1 million
unique short-lived prefixes.
We clarify that we excluded misconfiguration features related to
\emph{open SMTP mail relays} used in prior work~\cite{liu2015cloudy},
as we found these to be extremely
rare (only $\sim$1 in 10,000 SMTP relays are open).

These collected features are not directly linked to specific
vulnerabilities. Instead, the persistent presence of these technical
misconfigurations in an organization's network
serves as an indicator of poor security hygiene and highlights
shortcomings in the policies or controls designed for detecting and
remediating failures. These shortcomings, as shown in prior
work~\cite{liu2015cloudy}, consequently increases the potential risk
of successful cyberattacks.

\parheading{Malicious Activities}
Prior work~\cite{liu2015cloudy} relied on a small, hand-selected set of eleven
public blacklists.

Nearly a decade later, at least four of these lists are no longer maintained.

Therefore, we opted to use the Blacklist Aggregator (BLAG)~\cite{ramanathan20} as the
source of information about malicious activities.
As part of its operation, BLAG continuously
collects and archives the raw daily outputs of a large set of public
blacklists (157 sources, including previously three used by
Liu et al.~\cite{liu2015cloudy}) covering \emph{(1)} spam and abuse; \emph{(2)}
malware domains and hosts; and \emph{(3)} scanning or
attack activity. These blacklists date back to January 2019 and are
continuously updated.  We consume BLAG's daily snapshots
directly. In practice, a single day
typically contains more than 1 million IP addresses.
In our pipeline, when analyzing a certain organization in a given time
period, we assess whether any of the organization's IP addresses
appear on the BLAG lists.

\parheading{Data Pre-processing and Organization Mapping}
\label{sec:org_mapping}
While our forecasting targets are organizations, the network data
is linked to specific IP addresses of prefixes. To bridge this gap,
inspired by Liu et al.~\cite{liu2015cloudy}, we mapped global IP
addresses or address ranges to organizations using Regional Internet
Registry (RIR) data from ARIN~\cite{ARIN}, RIPE~\cite{RIPE},
APNIC~\cite{APNIC}, AFRINIC~\cite{AFRINIC} (LACNIC~\cite{LACNIC} was
excluded due to logistical constraints).
The data from these registries served two goals:
\emph{(1)} attributing IP addresses to victim organizations that
appeared in \cissm{}; and
\emph{(2)} identifying non-victim organizations that were later used
in our experiments (\secref{sec:evaluation:setup}).

    Due to the sheer size of our incident dataset compared to prior work's~\cite{liu2015cloudy}
    ($>$10$\times$ more incidents), manual mapping between organizations and RIR
    subnets was infeasible.
    Thus, we automated the mapping process by
    \emph{(1)} standardizing organizations' names in RIRs, as they tend to follow diverse
    naming conventions;
    \emph{(2)} clustering RIR subnets according to the semantic similarity between the
    standardized names; and
    \emph{(3)} mapping organizations from \cissm{} that suffered from security incidents
    to subnets according to semantic similarity between the organizations' names.
    We manually evaluated our conservative CISSM to RIR organization mapping on a random sample of 148 proposed organization matches in the RIR data. Of these, 133 were correct and 15 were incorrect, yielding an estimated precision score of 89.86\% (with 95\% Wilson CI~\cite{brown2001interval} of [0.84,0.94]).

\section{Forecasting System}
\label{sec:forecasting_system}

To enable cybersecurity incident prediction via geopolitical data, it
is necessary to extract appropriate representations that can be fed
into a predictive model.
One of the most powerful means for extracting such representation from
data is via \llm{}s~\cite{kenton2019bert,reimers2019sentence}.
Specifically, starting with pre-trained \llm{}s, their knowledge about
a certain field is typically extended using a given
corpus~\cite{lewis2020retrieval,gururangan2020don}. Subsequently, the representation they extract
based on this extended knowledge enable predictions. In this section,
we describe how we forecast cybersecurity incidents based on
textual geopolitical data alone, and how we fuse geopolitical data
with additional information to enhance prediction accuracy. Consistent with the framing in \secref{sec:threat_model}, our objective is predictive rather than explanatory: we aim to forecast whether an incident will occur through levaraging correlated information, not to identify causal factors of incidents.

\subsection{Forecasts Based on Textual Geopolitical Data}
\label{sec:geopolitical_forecast}
A popular paradigm for extending the knowledge of \llm{}s is
through retrieval~\cite{lewis2020retrieval}.
Here, the \llm{}'s context is augmented with relevant knowledge
retrieved from the corpus using a dedicated dense-retrieval
model~\cite{karpukhin2020dense,lewis2020retrieval}.
The augmented context is then fed to the model which produces a
representation that encodes the injected knowledge alongside the
remainder of data included in the prompt.

\begin{figure*}[t]
  \centering
  \includegraphics[width=0.8\textwidth]{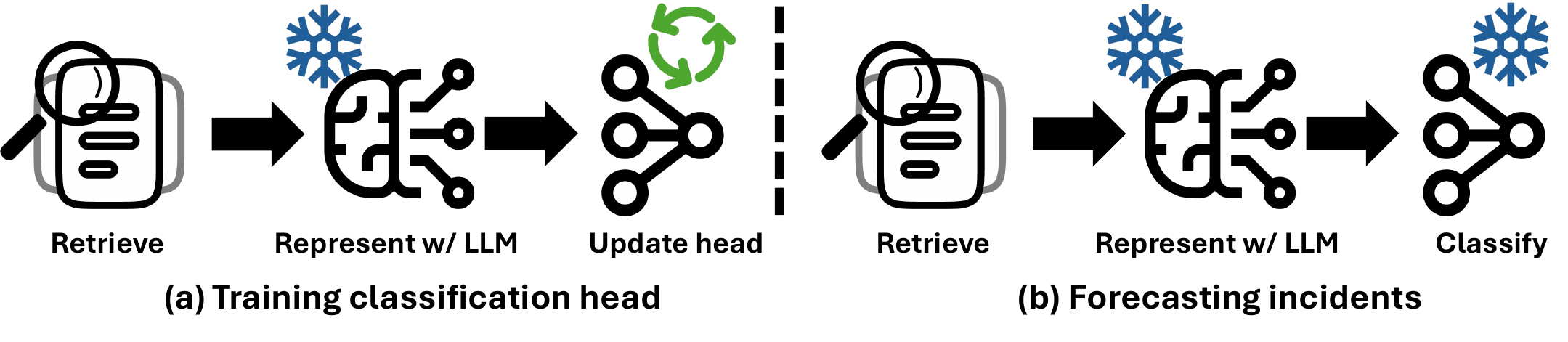}
  \caption{An illustrative overview of how we leverage retrieval to forecast
    cybersecurity incidents. (a) At time $t_\mathit{tr}$, we train a
    classification head to
    predict security incidents at time $[t_1, t_1+\tau]$, such that
    $t_1+\tau\le{}t_\mathit{tr}$ based on the
    representations produced by a frozen \llm{}.
    The \llm{}'s knowledge is augmented by feeding it relevant
    documents from time $\in[t_1-h, t_1]$ that we retrieve using
    pre-specified queries.
    (b) At test time, to predict security
    incidents
    at time $\in[t_2, t_2+\tau]$, such that $t_2\ge{}t_\mathit{tr}$,
    we produce representations using the
    same frozen \llm{}, through feeding it relevant documents
    retrieved from time $\in[t_2-h, t_2]$.
    The representations are then fed to a frozen classification head
    to estimate the likelihood of a cybersecurity incident.}
  \label{fig:overview:retrieval}
\end{figure*}

\figref{fig:overview:retrieval} provides an overview of how we
leverage retrieval to construct our forecasting system.
Given an organization for whom we aim to forecast attacks, we use a
retrieval model to find relevant information from the recent
history that can aid the prediction.
For instance, we may retrieve information about recent cyber attacks
against the organization, whether recent policies or economic
developments affected the organization, whether cyber attacks targeted
other organization in the said organization's country, etc., under the
hypothesis that such information is correlated with cyber incidents
that may target the organization in the near future.
Particularly, we perform the retrieval from our pre-processed corpus
of news articles and transcribed podcasts (\secref{sec:data_pipeline}).
We then aggregate the retrieved content and extend it with an
organization-specific prompt (e.g., ``Will ORG suffer from a
cyber incident in the following 30 days?''), and feed the resulting
text to the \llm{}.
Subsequently, the \llm{} produces a representation that we later feed
to a classification head trained to forecast attacks.

Another common approach for extending the knowledge of \llm{}s is
through fine-tuning on the original pre-training approach (e.g.,
next- or masked-token prediction)~\cite{gururangan2020don}. Specifically, instead
of providing relevant content in the input context, fine-tuning trains
the \llm{} on the knowledge corpus offline, and no retrieval is
conducted during deployment for augmenting the \llm{}s. Besides being
more computationally expensive than retrieval, prior
work also found that fine-tuning often falls short in terms of
performance~\cite{ovadia2024fine,kandpal2023large}.
Our preliminary experiments, too, showed favorable results with retrieval compared to fine-tuning.

\parheading{Document aggregation.}
For each query template, we retrieve the top-$k$ most relevant
documents from the indexed corpus after deduplication, yielding up
to $kq$ documents per organization-month across
the $q$ production query templates. (We set $k$=5 and $q$=6 to balance performance and efficiency, see \secref{sec:evaluation}.)
Each query's retrieved documents are prefixed by a short header naming the retrieval
intent (e.g., ``Documents about geopolitical conflicts of the
country'') and shuffled within their group. Randomizing within
each group discourages the classification head from overfitting
to positional artifacts.
The full concatenated input is then wrapped
in a system prompt that frames the model as a cybersecurity
analyst and lists the kinds of contextual signals to consider,
including geopolitical tensions, organizational changes,
infrastructure shifts, and indicators of cyber activity.
We dynamically truncate the document set to fit
within the 8,192 tokens context window for the input, trimming the lowest-ranked
documents so that the system prompt and the highest-ranked
retrieved documents are preserved.

\parheading{Query selection.}
Retrieval queries directly shapes the context of the \llm{} and resulting representation.
Starting with an initial set of queries hypothesized to retrieve information correlated with security incidents, we conducted a leave-one-out ablation in which each template was removed in turn and the model was retrained and re-evaluated on the validation set.
The ablation identified counter-productive queries that were removed, and left us with a final set that exhibited the best predictive accuracy on the validation set.
The final set of queries we use is listed in  \tabref{tab:queries} of \secref{sec:evaluation}.

\parheading{Temporal validity.}
We emphasize the importance of temporal validity when training and
testing the forecasting system~\cite{pendlebury2019tesseract}.
Suppose that a forecasting model is trained at time
$t_\mathit{tr}$. In this case, when training the model to forecast an
incident (or lack thereof) against an organization at time $t_1$, the
incident must have occurred at time $\in[t_1, t_1+\tau{}]$, where
$t_1+\tau\le{}t_\mathit{tr}$. Moreover, the model should only
be evaluated when forecasting incidents at times $\ge{}t_\mathit{tr}$
to avoid forecasting into the past.

\subsection{Forecasts Based on Fusion}
\label{sec:forecasting_system:fusion}

In addition to using textual geopolitical data alone to forecast
security incidents, we also explored means to incorporate other indicators to
enhance prediction accuracy. In particular, we studied means to
integrate data from
 GDELT (\secref{sec:gdelt}) and the network
measurements (\secref{sec:network_indicators})
alongside the geopolitical data into forecasts.
The former may
complement our geopolitical data by capturing general trends that may
not be reflected in the retrieved content fed to the \llm{}. The
latter may capture information related to the technical feasibility of
attacks that complement the geopolitical content that mainly captures
attacker motivation.

We considered two fusion strategies: input fusion and hybrid fusion~\cite{baltruvsaitis2018multimodal}. In the input-fusion setting, we enriched the textual prompt with the natural-language GDELT summary described in \secref{sec:gdelt}. Of the indicators extracted from GDELT, we used three in the prompt: \textit{material conflict events}, \textit{total geopolitical events}, and \textit{total media mentions}. As supporting evidence for their relevance, GDELT-based country-level mention volume showed noticeable association with monthly cyber incident counts in several countries; for example, the Pearson correlation between total mention volume and incident counts was $\rho=0.616$ in Ukraine and $\rho=0.558$ in Russia. While this pattern was not uniform across all countries, it suggests that GDELT-derived geopolitical context can provide a useful auxiliary signal.

In hybrid fusion, we combined input fusion with intermediate, feature-level fusion.
Specifically, we incorporated network features directly
into the classification head while aggregating it with the representation produced by
the \llm{} for the geopolitical and GDELT data.
To do so, we extract the \llm{} representation and concatenate it
with the feature vector extracted from network data, per
Liu et al.~\cite{liu2015cloudy}. The resulting joint representation is
then fed into the classification head.

Due to the dimensionality mismatch between the \llm{} representations and the network features (4,096 vs.\ 200 dimensions, respectively), we carefully concatenate the features so no single feature type dominates the classification outcome.
More specifically, we project the \llm{} representation and the network feature vector into a shared embedding space and normalized, to align their dimensionality and stabilize their relative scales. The two embeddings are then concatenated and passed through a learned gating mechanism: a sigmoid-activated linear layer produces a weight in $[0, 1]$ for each dimension of the concatenated representation, applied element-wise. While projection alone partially addresses the scale mismatch, the gate additionally allows the model to suppress individual noisy dimensions from either modality. The gate is initialized with a positive bias so that it initially incorporates all dimensions and subsequently learns which dimensions to down-weight. The resulting gated representation is then fed into a multi-layer perceptron that produces the final classification.

\section{Limitations}
\label{sec:limits}

Several limitations should be taken into account when assessing our
forecasting system and interpreting our results.
First, some organizations may fail to report or may underreport cybersecurity incidents~\cite{amir2018firms,romanosky2016examining} and certain
incidents may not be captured by \cissm or \csidb.
This may stem, among others, from lacking policies mandating the reporting of
cybersecurity incidents~\cite{schmitz2023defining}, or inherent biases in
cybersecurity incident repositories.
Consequently, the recall
(i.e., true positive rates) of our system may be underestimated, while the
false positive rates may be
overestimated. Moreover, the training data we use may potentially be
noisy, thus harming accuracy. In other words, more complete and precise incident
reporting is expected to lead to improved system performance. As all
methods we test are equally affected by potential reporting issues, we have no reasons to believe that our
conclusions regarding their relative performance would be affected if the issues are resolved.

Second, our manual assessment showed that merging from \csidb{} and
\cissm{} has improved incident dates' accuracy (\secref{sec:data:incident}).
Nonetheless, $\sim$8\% of audited
records retained inaccurate dates, potentially shifting incidents across
the boundary of our prediction window and harming evaluation precision.
It was infeasible to manually update the dates of thousands of incident
records, and other automated approaches we tested (e.g., using
conversational agents) did not improve dates' accuracy.
Moreover, even if it was feasible to trace every incident to all relevant public disclosures, reports frequently fail to disclose the actual date that the cyber incident began. This discrepancy may distorts the real timeline of events and impacts the labels we use.
We expect that making the dates more accurate
(e.g., via improved collection methodology in \cissm{} or \csidb{})
would also help make the evaluation results more precise.

Third, while we did our best to precisely map organizations to
network indicators (\secref{sec:org_mapping}), some imprecisions
persist. Prior work conducted manual mapping, which remained feasible
with the relatively small dataset considered (roughly 700 incidents
and organizations in Liu et al.~\cite{liu2015cloudy}). In contrast, we
consider a significantly larger dataset (more than 10$\times$ the size of Liu et al.'s),
precluding manual mapping. However, our manual audit of a random
sample of mappings showed a high precision of $\sim$90\%
(\secref{sec:org_mapping}). Thus, we expect that our evaluation results
of models leveraging network-based indicators would remain mostly
unchanged even with a more accurate mapping procedure.

Fourth, our reliance on geopolitical news introduces a set of limitations that are distinct from those that affect previous work based on network indicators. Specifically, some prominent organizations and sectors may attract more media attention and coverage, causing them to be over-represented in the data compared to smaller entities with less public visibility. This coverage imbalance can bias the forecasts, potentially making them more reliable for well-covered targets than for those operating outside the spotlight. Consequently, if the underlying dataset lacks coverage of certain regions or organizations entirely, the system cannot be expected to perform well for those entities. This limitation closely parallels the mapping challenges inherent in network-based prediction, where an organization's specific infrastructure is indistinguishable from others sharing the same cloud services.

Lastly, because our input is public discourse, our system inherits
the vulnerabilities of that discourse to misinformation, coordinated
narrative manipulation, and adversarial framing. An adversary aware
of our pipeline could, in principle, attempt to poison the context
by seeding misleading articles. To address this, the user may narrow the list of sources of geopolitical data to reputable ones that may be challenging to manipulate. Indeed, we find that restricting the sources to the top-10K domains yields to only a minor drop in the overall forecasting accuracy (\secref{sec:evaluation:results}).

\section{Evaluation} \label{sec:evaluation}

\subsection{Experimental Setup}
\label{sec:evaluation:setup}

\parheading{Models}
As an \llm{}, we used \texttt{LLaMA-2-7B}~\cite{touvron2023llama} due to its
relatively early cut-off date (September
2022) and the superior accuracy it
enabled compared to other models we tested (e.g., Longformer~\cite{beltagy2020longformer})
. We extracted representations from an intermediate
layer (specifically, 23 of 32), as it led to better forecasts than later
layers
. Subsequently, we fed the representations to a multilayer perceptron (MLP)
that we used as a
classification head (three hidden layers with 256 neurons each
followed by GELU activations)
trained via binary cross-entropy.

\parheading{Retrieval Configurations}
For retrieval, we used the queries six listed in
\tabref{tab:queries}. These queries, as hypothesized
earlier (\secref{sec:geopolitical_forecast}), led to high forecast accuracy. For each
query, when predicting incidents at time $t$, we extracted most
relevant five passages from up to $h$=9 months
prior to $t$ (selected per hyperparameter search). As noted before (\secref{sec:data_ingestion}), we used the
\texttt{Arctic-embed-m-v2.0} retrieval model for best
performance.

\begin{table*}[t]
  \caption{\label{tab:queries}The queries used to
      retrieve relevant geopolitical data to enable forecasts.}
  \centering
  \begin{tabular}{l p{0.92\textwidth}}
    \toprule
    Q1: & ``Are ORG and similar companies at risk of
    cyber attacks?'' \\
    Q2: & ``What are major economic or political developments in
    ORGCOUNTRY that could affect businesses?'' \\
    Q3: & ``What regulations or policy changes have recently impacted
    ORG in ORGCOUNTRY?'' \\
    Q4: & ``Has ORG been involved in any controversies,
    leadership changes, or restructuring?'' \\
    Q5: & ``Are there known conflicts between ORGCOUNTRY and other
    nations that might influence cyber or economic targeting?'' \\
    Q6: & ``Has ORG received significant public or media
    attention recently that could attract adversarial interest?''\\
    \bottomrule
\end{tabular}
\end{table*}

\parheading{Metrics}
We adopted the metrics used by Liu et al.~\cite{liu2015cloudy}
for evaluation.
Specifically, we computed receiver operating characteristic (ROC)
curves, which report true positive rates (TPRs) and false positive
rates (FPRs) at different operating points.
The higher are the TPR and the area under the curve (AUC) and the
lower is the FPR, the better is the forecast accuracy.

\parheading{Data Splits}
\label{sec:data_split}
We derived the victim organizations and incident dates from our incident dataset  (\secref{sec:data:incident}). Unless stated otherwise,
for each victim, we set the prediction date to be within $\tau$=30 days before the actual incident (selected uniformly at random). We focused on this 30-day window to prioritize immediate tactical warning signals over long-term strategic forecasting.
However, we also tested prediction up to $\tau$=1 year into the future.
Both prediction horizons ($\tau$s) we considered are consistent with prior work~\cite{liu2015cloudy}.

    For non-victims, to mitigate model reliance on spurious static features like the organizations' names, we implemented a two-stage sampling strategy. Starting with non-victim organizations sampled from the Regional Internet Registry (RIR) records, we identified that a simple linear classifier based on 1-grams could distinguish victims from non-victims based solely on naming artifacts (ROC AUC $>$0.5). Therefore, we prioritized non-victim organizations RIR records that the 1-gram classifier (incorrectly) predicted as victims. Subsequently, we augmented the set of non-victims with organizations that suffered from incidents at some point; however, we were careful to sample these organizations at least one year away from the reported incident date to account for potential reporting delays. Our sampling approach forced the model to ignore spurious characteristics, such as organization names, and instead focus on geopolitical changes that indicate potential security incidents.

Overall, we ran our evaluation on 16,954 records, with 7,120 pertaining
to victim organizations (i.e., ones who suffered from an incident at a given
time) and 9,834 belonging to non-victim organizations (i.e., ones who
did not suffer from an incident at a given time).
For better reliability, we split the data at random into five folds, with $\sim$80\% of
samples used for training and the test samples selected from the remaining $\sim$20\%,
and averaged our results across folds.
To ensure temporal validity, as discussed
previously (\secref{sec:geopolitical_forecast}), when testing predictions at time
$t$, we avoided training our models on any data
(dependent or independent variables) from a later time $>t$.
Hence, we gradually trained models till particular target dates, and
tested them on the $\tau$=30 days immediately following these
dates. Moreover, our test samples were taken only from 2024 and 2025,
after the cut-off date of the \llm{} we used.

\parheading{Baseline}
As a baseline method to compare our models against, we implemented Liu
et al.'s~\cite{liu2015cloudy} forecasting model based on network
indicators.
Specifically, we implemented a Random Forests (RF)
classifier that takes misconfiguration and malicious
activity features as input and predicts the likelihood of attacks.
Besides optimizing the hyperparameters of the RF model, we further improved
the performance of the model using an MLP for prediction.
The comparison with this baselines serves two
goals. Primarily, it shows where our approach stands compared to the
state-of-the-art cybersecurity incident forecasting method.
Second, it helps assess how geopolitical data (mostly encoding
adversary motives) compares to network indicators (mostly indicating
organizations' security hygiene) when predicting attacks.

\subsection{Experimental Results}
\label{sec:evaluation:results}

\begin{figure}[t]
  \centering
  \begin{subfigure}{0.43\columnwidth}
    \centering
    \includegraphics[width=\linewidth]{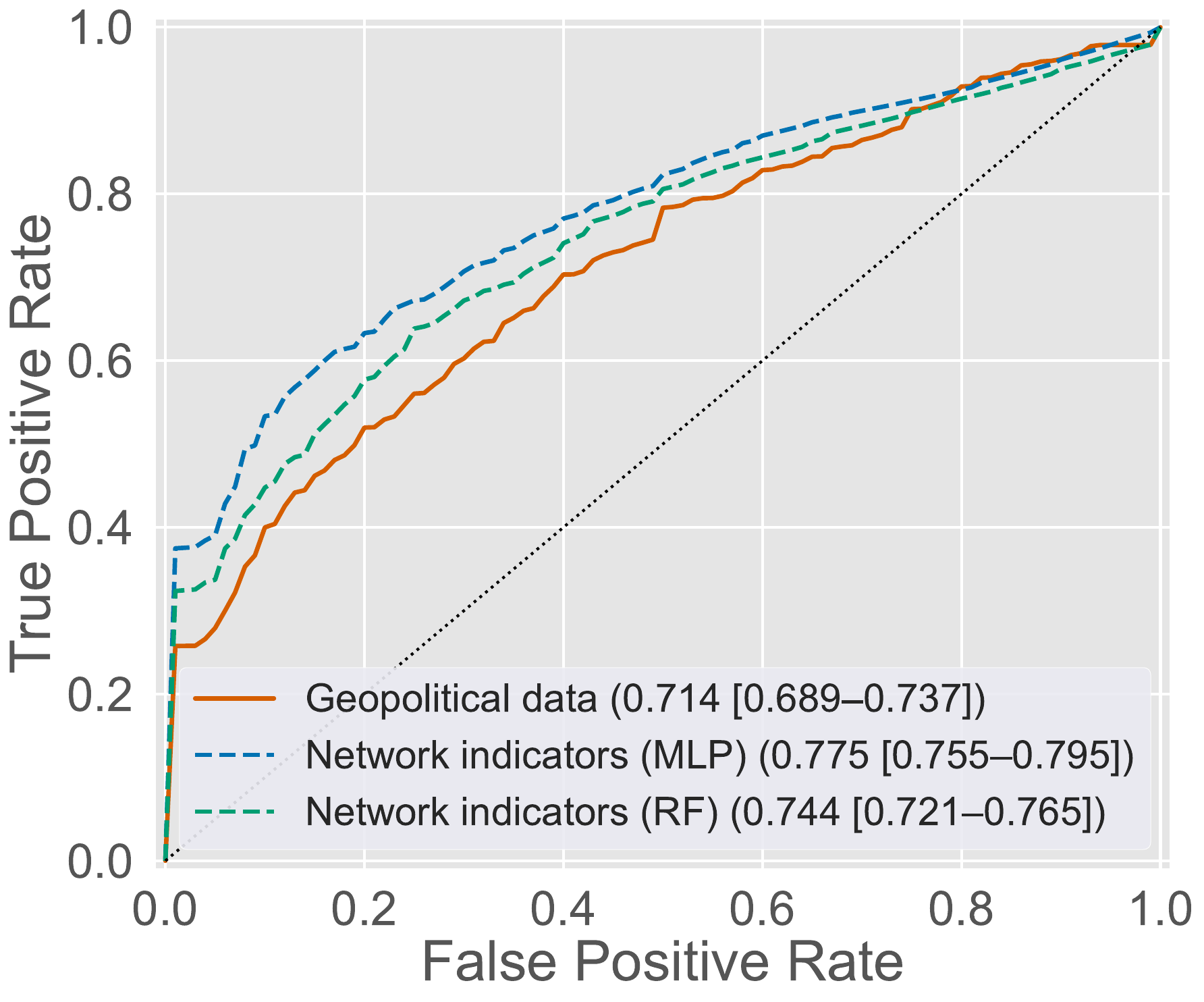}
    \caption{Geopolitical vs.\ network data}
    \label{fig:simple_roc}
  \end{subfigure}
  \hfill
  \begin{subfigure}{0.43\columnwidth}
    \centering
    \includegraphics[width=\linewidth]{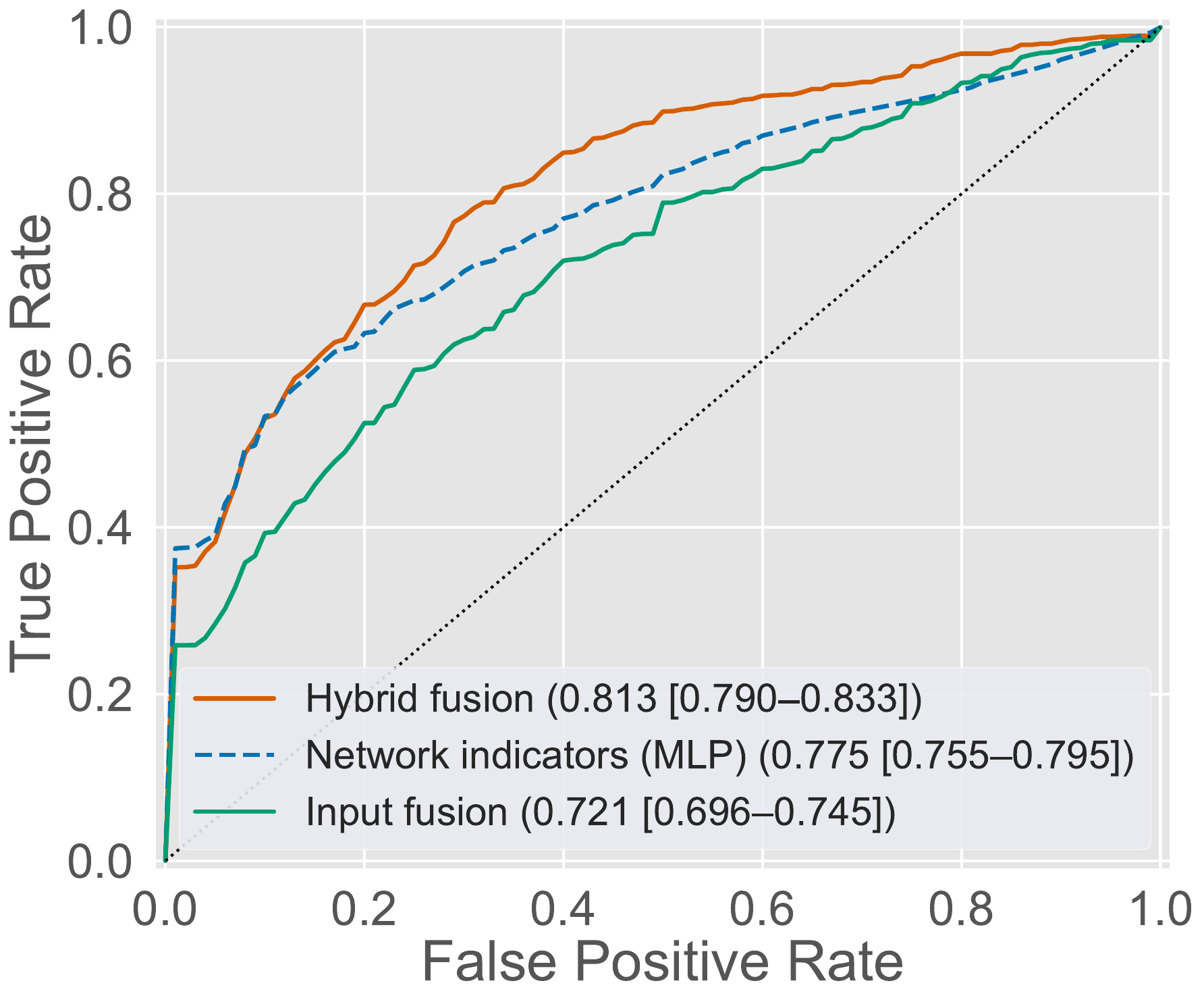}
    \caption{Fusion-based approaches}
    \label{fig:fusion_roc}
  \end{subfigure}
  \caption{ROC-curve comparisons:
  (a) presents the performance of our approach using geopolitical data alone
  against the baseline based on network indicators, while
  (b) shows the performance of fusion-based forecasting. ROC AUCs and their 95\% CIs are reported in the legend.}
  \label{fig:combined_roc}
\end{figure}

\parheading{Forecasting Without Fusion}
\figref{fig:simple_roc} shows the forecasting accuracy of our
\llm{}-based approach using geopolitical data and compares it to
baselines relying on network features.
As perhaps expected for short-term predictions ($\tau$=30 days),
network indicators that capture misconfigurations and malicious activity
from orgniazations' networks achieve higher accuracy (74.4\% ROC AUC for the
original RF-based model, and 77.5\% AUC after our improvement with an MLP).
Nonetheless, our approach attains competitive performance, with 71.4\% AUC and 95\% confidence intervals (CIs) substantially overlapping those of the RF model,
confirming the viability of geopolitical data that helps capture the
adversary's motivation as a source of information for predicting cybersecurity
incidents.

\parheading{Fusion Leads to Best Performance}
Notably, fusing geopolitical data and network indicators  leads to the
best forecasting accuracy (see \figref{fig:fusion_roc}).
Input fusion with GDELT features slightly pushes the performance of our
system (from 71.4\% to 72.1\% AUC). However, the improvement is not statistically significant.
The most pronounced improvement is achieved via hybrid fusion, with
GDELT features in the input, and network features in an intermediate phase,
leading to $\sim$5\% higher ROC AUC compared to the baseline based on an
MLP (81.3\% vs.\ 77.5\% AUC with negligble overlap between the 95\% CIs).
This result shows that combining feature types that encode information about
both adversary motives and organizations' security hygiene yields substantially
better forecasts.

\begin{figure}[t!]
  \begin{center}
    \includegraphics[width=0.45\textwidth]{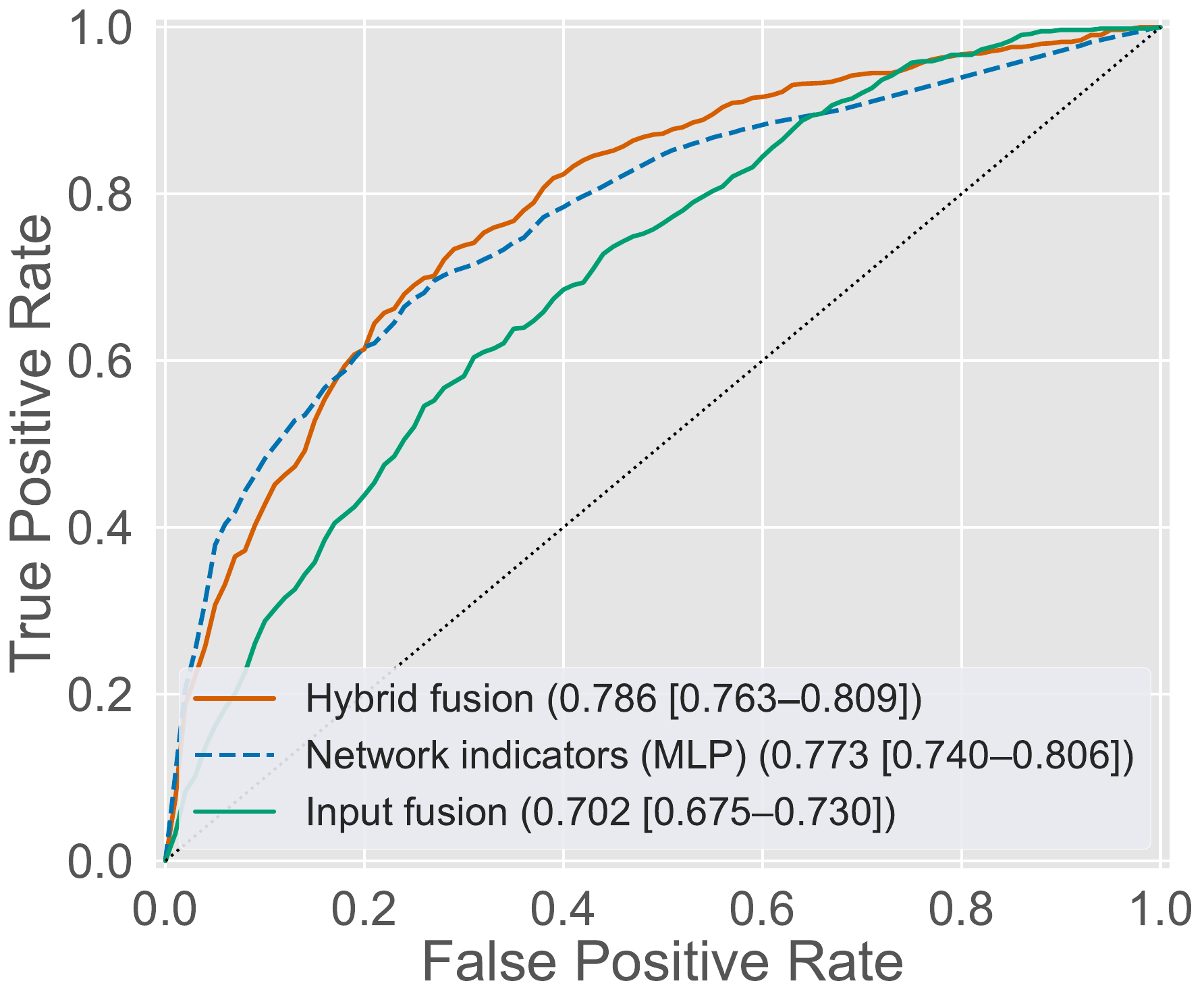}
  \end{center}
  \caption{Predicting incidents 1 year into the future.}
  \label{fig:long_term_roc}
\end{figure}

\parheading{Predictions Into Longer Horizons Are Possible}
Using our data splits, we also evaluated predictions
$\tau$=1 year into the future.
In particular, we tested
whether models trained until
2023 or 2024 can predict incidents that have occurred during
these years.
MLP based on network indicators alone attains strong results, outperforming the
\figref{fig:long_term_roc} presents the results. Here too, we find that the
model fusing geopolitical and GDELT data at the input (77.3\% vs.\ 70.2\% AUC).
However, hybrid fusion, combining both geopolitical and network data provides
the most accurate forecasts (78.6\% AUC).

\begin{figure}[t]
  \centering
  \begin{subfigure}{4.5cm}
    \centering
    \includegraphics[width=\linewidth]{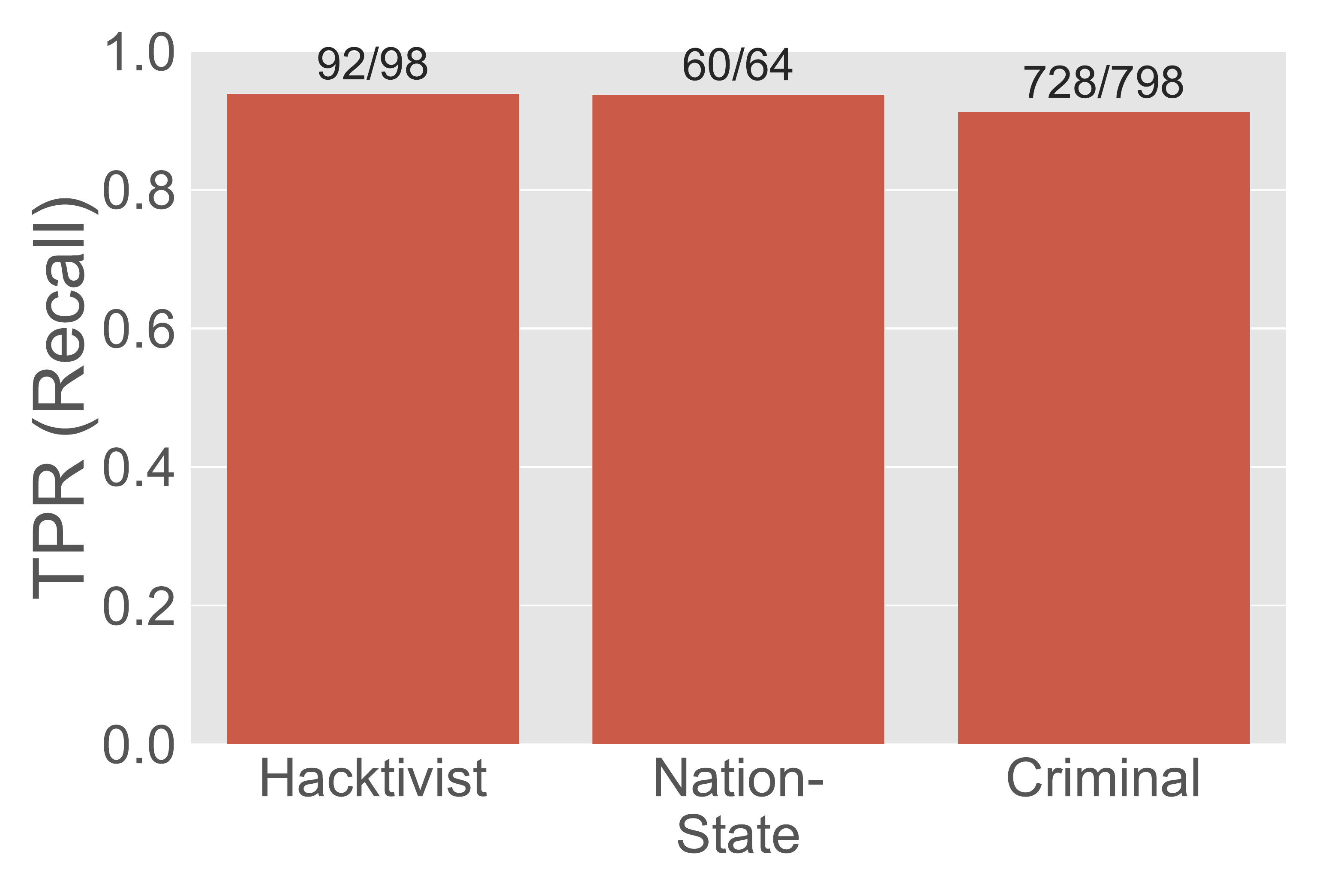}
    \caption{Actor type}
    \label{fig:adv_type_tprs}
  \end{subfigure}
  \begin{subfigure}{4.5cm}
    \centering
    \includegraphics[width=\linewidth]{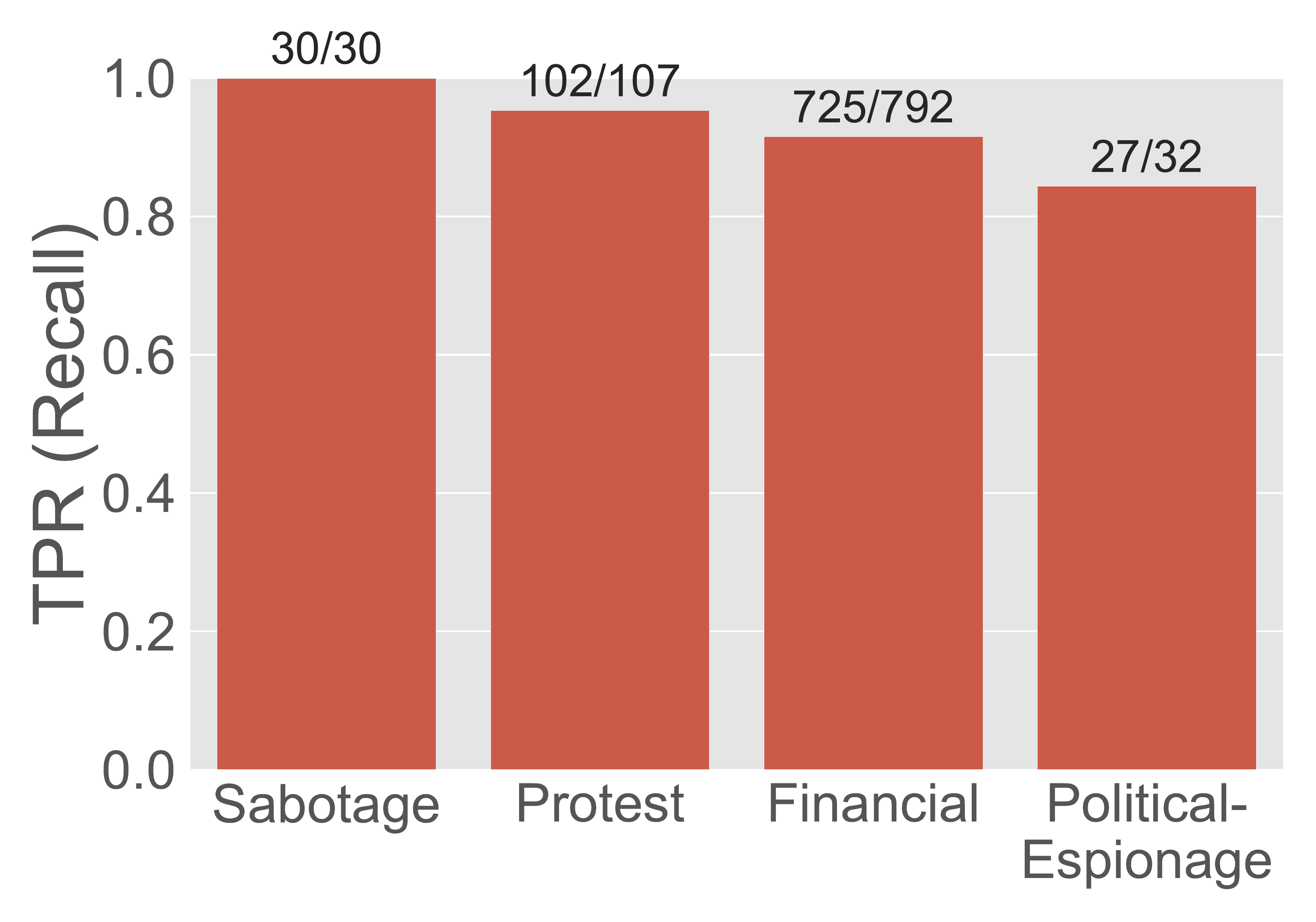}
    \caption{Attack motive}
    \label{fig:attack_motives_tprs}
  \end{subfigure}
  \hfill
  \caption{Forecasting TPRs (at max F1 score) of our
    approach using geopolitical data alone for incidents involving
    different adversary (a) types (b) and motives.
    }
  \label{fig:cat_tprs}
\end{figure}

\parheading{When Is Geopolitical Data Most Helpful?}
To uncover when geopolitical data is most helpful,
we analyzed how forecasting accuracy varies across adversary types,
adversary  motives, and organization sectors, as reported in \cissm{}.
For each of these, we measured and reported the TPR of our model
based on  geopolitical data alone where the F1-score is maximal (\figref{fig:cat_tprs}).
The results show intuitive changes across categories.
For adversary types,
we find that the
prediction accuracy is higher for nation-states and hacktivists (that
may be more driven by geopolitical events) than for criminals (who may
be more opportunistic).
Along the same lines, for adversary motives,
we find that attacks motivated by sabotage and protest are more accurately
predicted than other attacks (e.g., financially motivated ones).
Finally, there is less variance across sectors, yet, we find that incidents against
organizations in healthcare and education are predicted with slightly higher accuracy
than incidents targeting other industries.

\parheading{What Makes Our Approach Work?}
To better understand what makes our approach succeed, we analyzed how the context
affects prediction accuracy when using geopolitical data alone.
Primarily, we ablated the queries used to for retrieving documents to form the
context to our model (\tabref{tab:queries}) and measured how excluding each query
affects the ROC AUC. We found that Q1 (retrieving text related to the
cyber risk of organizations) and Q3 (finding documents on country-level regulations
that may impact organizations) were most critical--excluding led to a 4.2\%
AUC drop.

\begin{figure}[tb]
  \centering
  \includegraphics[width=\columnwidth]{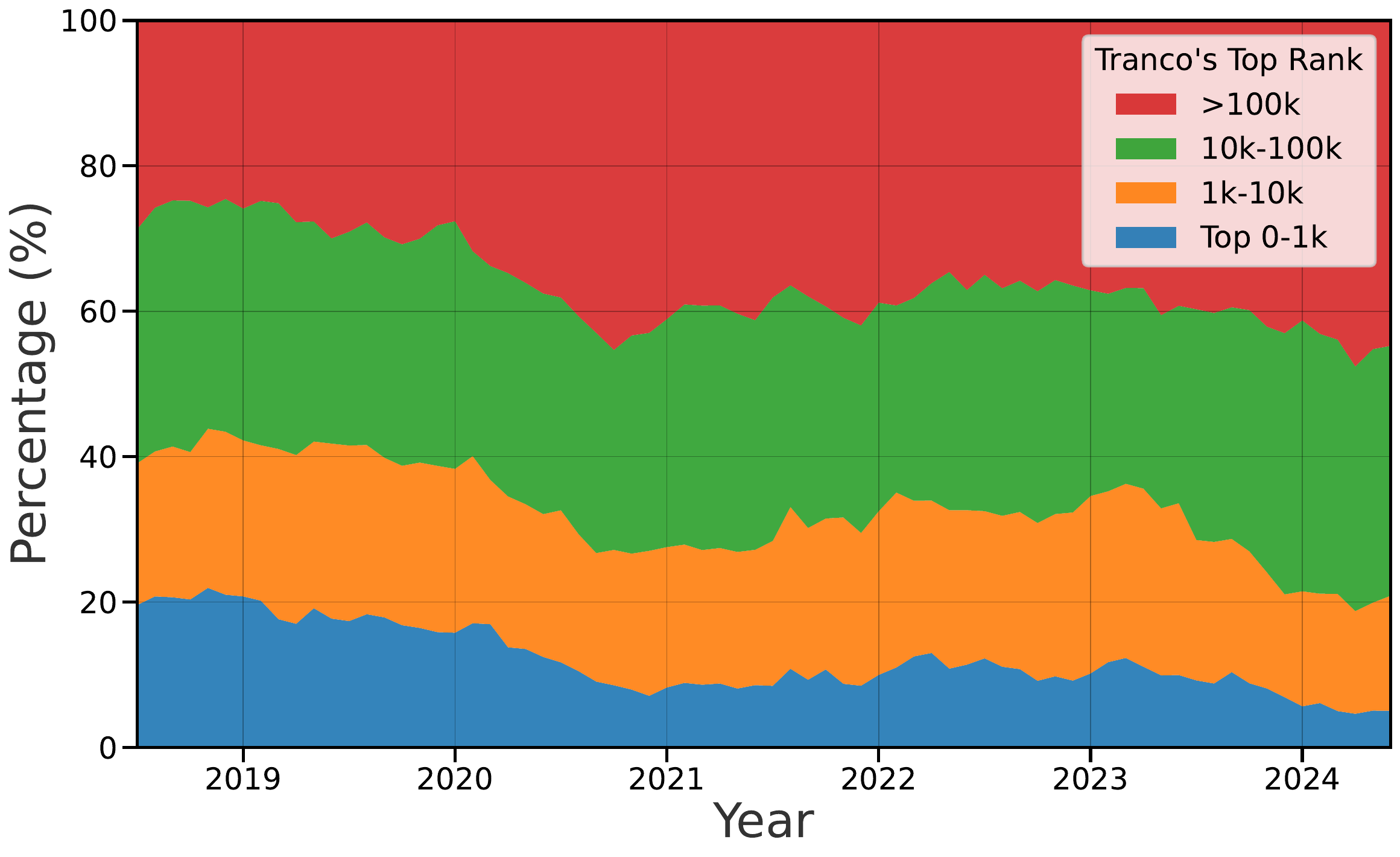}
  \caption{The fraction of retrieved documents categorized by their Tranco rank.}
  \label{fig:org_context_sources}
\end{figure}
Furthermore, we explored what sources of geopolitical news were played a
central role in forming the context fed to our \llm{} (\figref{fig:org_context_sources}).
We found that popular websites ranked part of the top 10,000 Tranco\footnote{List available at \url{https://tranco-list.eu/list/4N6KX} (accessed 17 Dec 2025).} domains \cite{LePochat2019}
consistently contributed almost 20--40\% of the retrieved articles selected to build
the \llm{}'s ingested context.
Interestingly, this is significantly higher than the overall rate of such articles in
the general corpus we collected: roughly only 1--5\% of articles were from websites in
the top 10,000.
In other words, articles from popular domains (e.g., Yahoo News) played a vital role
in enabling predictions despite being relatively rare.
This indicates that defenders may not need to rely on less reputable sources to
forecast potential cyber incidents.
Indeed, we find that basing forecasts only on data retrieved from top-10K domains has negligible effect on model's accuracy, leading to a minor drop in our models accuracy (from 71.4\% to 69.6\% AUC when no fusion is used).

\section{Conclusion}
\label{sec:conclusion}

We presented a framework for forecasting cybersecurity incidents using geopolitical data mined from public news articles and podcasts. By retrieving organization-relevant context and encoding it with an \llm{}, our approach produces useful predictive representations from unstructured text, achieving competitive performance even when used alone. Additionally, our results show that such data complements traditional network-based indicators where hybrid fusion of geopolitical and network features yielded the strongest performance improving ROC AUC from 77.5\% with network indicators alone to 81.3\%. Our analysis further suggests that geopolitical data is especially informative for incidents tied more closely to broader events and motives, such as protest- and sabotage-related attacks.  Overall, these findings highlight the value of combining independent sources of evidence for cyber-risk forecasting and suggest that public geopolitical information can strengthen practical forecasting systems used for proactive defense and risk assessment.

\bibliographystyle{ACM-Reference-Format}
\bibliography{references}

\appendix
    \section{Model Accuracy Across Years}
    \label{app:year-comparison}

    To verify that the model does not rely on memorized knowledge from its pre-training phase (data leakage), we evaluated performance across years. If the model were depending on the LLM's internal knowledge base (cutoff date of early 2023), we would expect a significant performance degradation on subsequent unseen data. Instead, we observe robust temporal generalization, with AUC scores increasing in the post-cutoff period, peaking in 2025.

    \begin{figure}[H]
        \centering
        \includegraphics[width=0.95\linewidth]{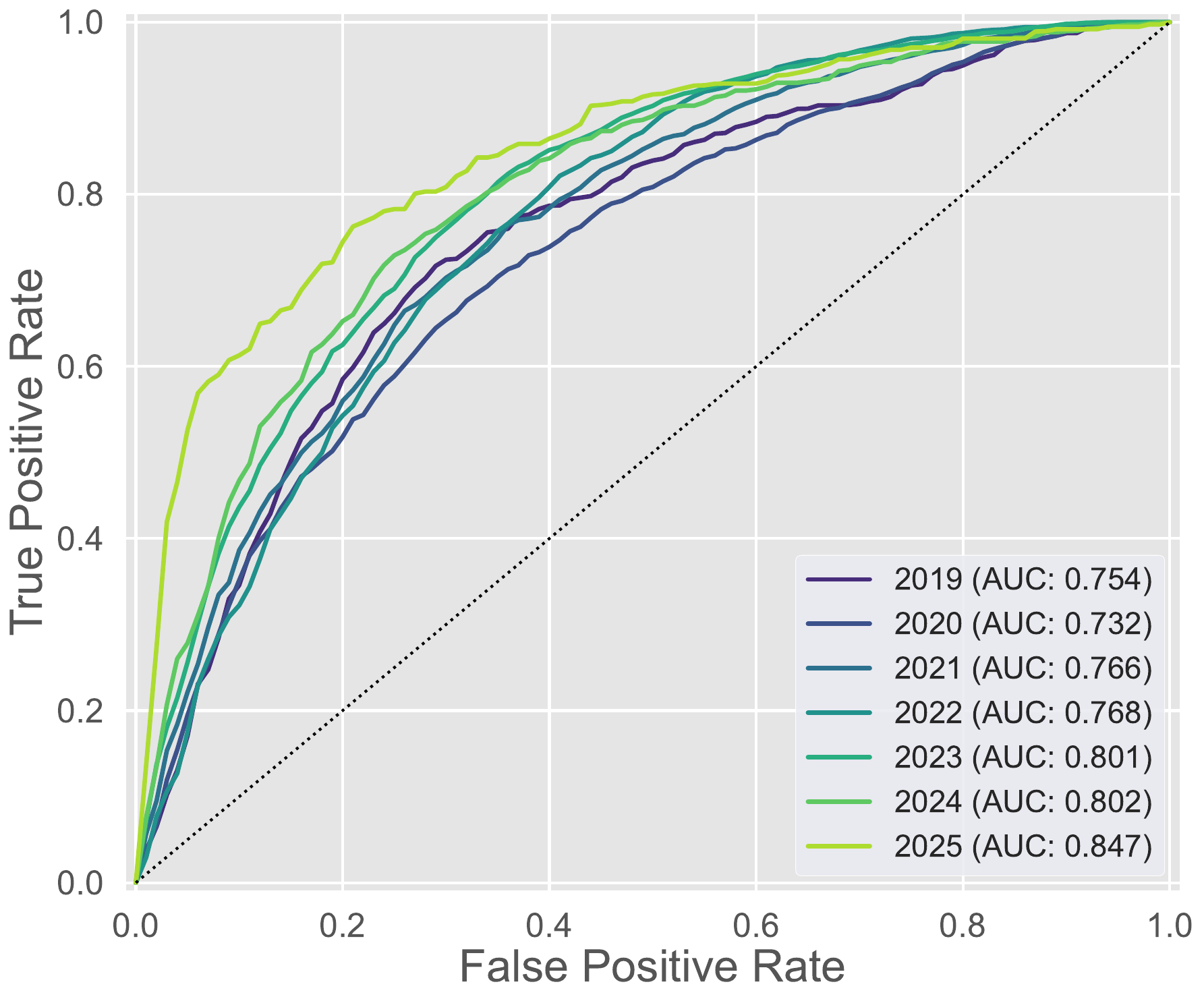}
        \caption{Model performance across years 07-2019 to 05-2025}
        \label{fig:year-comparison-val}
    \end{figure}

\section{Intermediate-Layer Representation Comparison}
\label{app:layer-comparison}

We evaluated pooling representations across different depths of the LLaMA-2-7B model. While performance remained robust across the network, it peaked at Layer 23 (AUC of 0.724), marginally outperforming both shallower layers and the final layers. This aligns with prior findings that deep intermediate layers often capture the richest semantic features, whereas the final layers become increasingly specialized for the next-token prediction objective \cite{skean2025layer}. This motivates our selection of Layer 23 as the optimal representation.

\begin{figure}[h]
    \centering
    \includegraphics[width=0.9\linewidth]{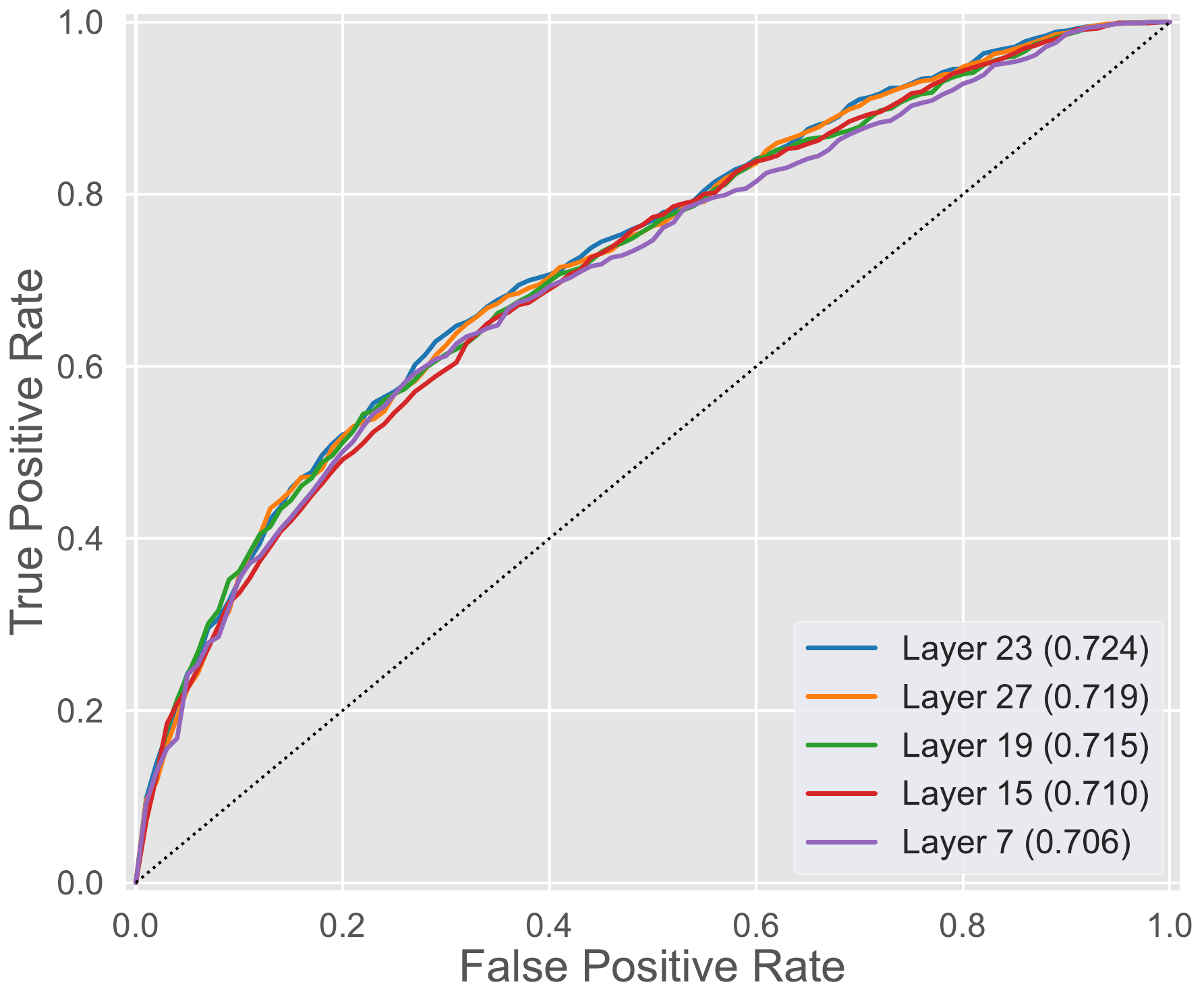}
    \caption{Performance comparison across different LLaMA-2-7B
    layers. We found intermediate layers to capture semantic
    better suited for cybersecurity incident forecasting than early or
    late layers.}
    \label{fig:layer-comparison}
\end{figure}

    \section{Name-only Classification Analysis}
    \label{app:name-only}

    To better understand whether organization names alone carry enough
    predictive signals, we trained a simple 1-gram logistic regression
    classifier on a sample of victims and non-victims. For victims, we
    used organizations with at least one confirmed incident in our
    dataset. For non-victims, we selected organizations that had no
    recorded incidents, using organization names obtained from
    Regional Internet Registry (RIR) datasets.

    This classifier achieved strong separability (AUC~0.815), but the
    signal is largely explained by biases in the underlying
    data. Organization names in \cissm are human-written and tend to
    reflect sectors that are more frequently targeted or more
    frequently reported, such as hospitals, universities, and public
    institutions—hence the prominence of terms
    like \emph{health}, \emph{hospital}, or \emph{university}.

    On the other hand, many of the negative coefficients correspond to
    patterns that appear in our RIR-derived non-victim sample. Terms
    such as \emph{beijing} and \emph{zhejiang} often occur in Chinese
    organization names that rarely appear in \cissm, probably
    reflecting differences in reporting or disclosure
    practices. Similarly, words like \emph{computer}, \emph{cable},
    or \emph{ibm} are common in registered or infrastructure-related
    organization names found in RIR datasets, which rarely appear
    among reported victims. Table~\ref{tab:1gram-features} lists the
    highest positive and negative coefficients, illustrating these
    naming and reporting biases.

    \begin{table}[H]
        \centering \caption{Top 1-gram features for victims (positive
        coefficients) and non-victims (negative
        coefficients).}  \label{tab:1gram-features}

        \begin{tabular}{l r  @{\hskip 1.5cm}  l
            r} \toprule \multicolumn{2}{c}{\textbf{Victim-associated
            features}}  & \multicolumn{2}{c}{\textbf{Non-victim
            features}} \\ \cmidrule(r){1-2} \cmidrule(l){3-4} \textbf{Feature}
            & \textbf{Coef.}  & \textbf{Feature}
            & \textbf{Coef.} \\ \midrule health        & 3.711  &
            beijing      & -3.320 \\ company       & 3.260  & computer
            & -2.104 \\ healthcare    & 3.232  & stichting    &
            -2.041 \\ inc           & 3.185  & kommune      &
            -1.908 \\ corporation   & 3.182  & womens       &
            -1.811 \\ hospital      & 3.059  & polska       &
            -1.780 \\ com           & 2.605  & publishing   &
            -1.776 \\ university    & 2.390  & gemeente     &
            -1.750 \\ co            & 2.254  & trading      &
            -1.735 \\ district      & 2.254  & harbin       &
            -1.711 \\ public        & 2.062  & zhejiang     &
            -1.636 \\ llc           & 2.062  & acc          &
            -1.633 \\ ltd           & 2.048  & industries   &
            -1.620 \\ ministry      & 2.008  & italia       &
            -1.589 \\ holdings      & 1.912  & nederland    &
            -1.560 \\ parliament    & 1.838  & construction &
            -1.536 \\ comune        & 1.828  & ibm          &
            -1.536 \\ llp           & 1.695  & chemical     &
            -1.526 \\ ukrainian     & 1.648  & cable        &
            -1.513 \\ municipality  & 1.645  & properties   &
            -1.492 \\ \bottomrule \end{tabular}
    \end{table}

    \begin{figure}[H]
        \centering \includegraphics[width=0.9\linewidth]{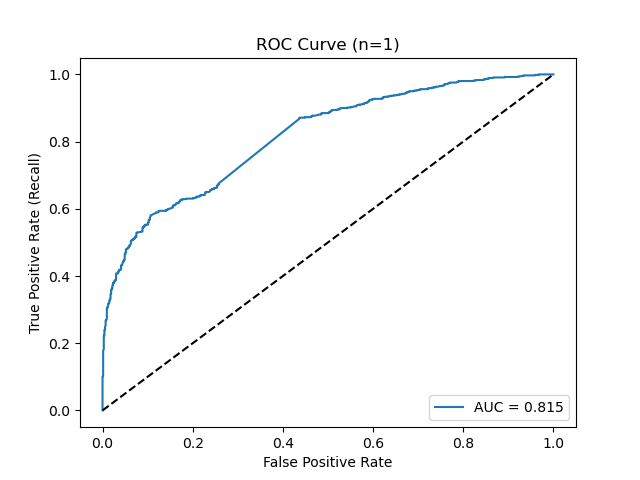} \caption{1-gram
        classifier performance over organization
        names-only} \label{fig:name-only}
    \end{figure}

    As shown, words associated with public-sector institutions
    (e.g., \emph{health}, \emph{hospital}, \emph{ministry}, \emph{municipality})
    tend to correlate with victimhood, whereas geographic or
    infrastructure-related terms
    (e.g., \emph{beijing}, \emph{publishing}, \emph{construction}) are
    more common among non-victims. This highlights how naming and
    reporting biases in the underlying data can create artificial
    separability, which in turn requires us to interpret such findings
    with caution. This requires us to explicitly handle such biases so
    they do not appear in downstream results as false signals, for
    example by ensuring that non-victim organization names more
    closely resemble those appearing in the victim set.

    \section{Correcting CISSM}
    \label{app:correcting_cissm}

    We evaluate the effectiveness of our date-correction layer by conducting a manual analysis on a random sample of incidents from the CISSM dataset. Specifically, we randomly sampled 50 incidents, uniformly across different years, and compared the date accuracy of: (i) the original CISSM dataset, and (ii) CISSM augmented with the corrective layer based on CSIDB (i.e., using the CSIDB date when it is earlier than the CISSM date).

    Each sampled incident was manually examined by one of the authors. For every incident, we opened the original source link in the dataset and, when necessary, consulted additional independent news sources. The incident date was determined based on the information in the article(s). In two cases, the original sources pointed to Telegram channels that were no longer accessible; these incidents were excluded from the evaluation. The final evaluation set therefore consists of 48 incidents.

    Among these 48 incidents, 15 incidents (31.25\%) appear only in CISSM, while 33 incidents (68.75\%) appear in both CISSM and CSIDB. On this evaluation set, the original CISSM dataset correctly identified the incident date for 37 out of 48 incidents, corresponding to an accuracy of 77.08\%. After applying the corrective layer---that is, replacing the CISSM date with the CSIDB date when the latter is earlier---the accuracy increases to 91.67\% (44/48 correct dates). When restricting the evaluation to the intersection of CISSM and CSIDB only (33 incidents), the accuracy further increases to 96.96\% (32/33 correct dates), indicating that the corrective layer is particularly effective when both datasets report the same incident.
    \noindent We also observe that, for incidents from January 2025 onward, the CISSM dataset introduces two separate fields: a \emph{published date} and an \emph{incident date}. This suggests that the CISSM team has recently identified and started addressing the date-misattribution issue internally. However, earlier incidents in the dataset remain uncorrected and continue to exhibit systematic date inconsistencies.
    Table~\ref{tab:cissm_correction} summarises the evaluation set and the resulting accuracy figures.

    \begin{table}[t]
        \centering
        \caption{Summary of the manual evaluation of CISSM vs.\ CISSM with the CSIDB-based corrective layer over 48 incidents.}
        \label{tab:cissm_correction}
        \begin{tabular}{lrr}
            \hline
            Category & Count & Percentage \\
            \hline
            Incidents only in CISSM & 15 & 31.25\% \\
            Incidents in CISSM $\cap$ CSIDB & 33 & 68.75\% \\
            Correct dates in CISSM & 37 & 77.08\% \\
            Correct dates after correction & 44 & 91.67\% \\
            Accuracy on intersection only & 32 & 96.96\% \\
            \hline
        \end{tabular}
    \end{table}

\begin{figure}[t]
        \centering
        \includegraphics[width=\linewidth, page=1]{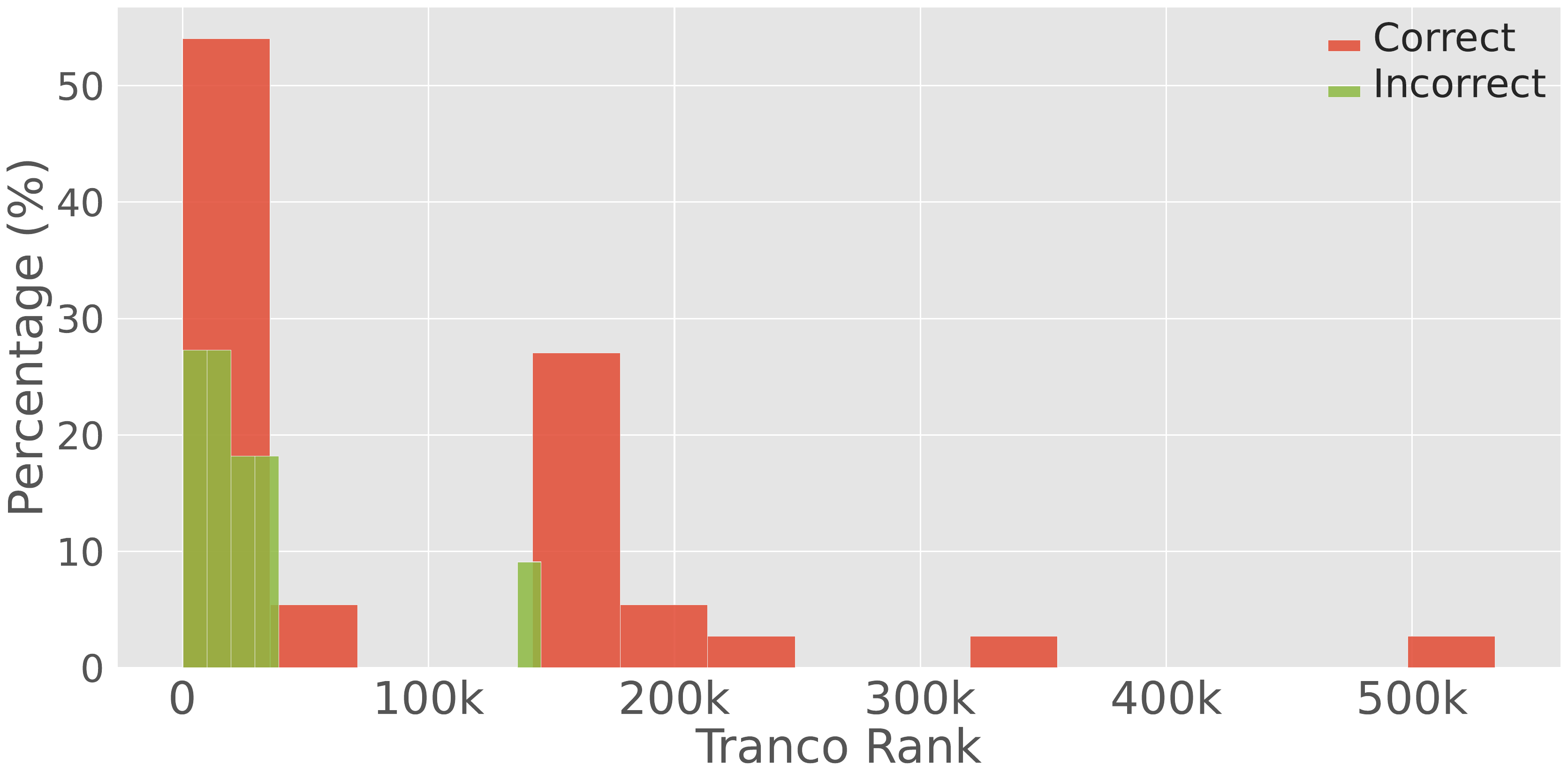}
        \caption{Distribution of Tranco ranks for sources associated with correctly and incorrectly dated CISSM incidents. Errors are observed for both high- and low-ranked domains, indicating that source reputation alone does not explain the date inconsistencies.}
        \label{fig:tranco_cissm}
    \end{figure}

    \noindent To further investigate whether CISSM date errors are correlated with source quality, we examined the Tranco~\cite{tranco} rank distribution of domains associated with correctly and incorrectly dated incidents. Figure~\ref{fig:tranco_cissm} shows the distribution of Tranco ranks for these two classes. We observe that incorrect dates occur across the full spectrum of Tranco ranks, including highly ranked and reputable news outlets. This suggests that CISSM date inaccuracies are not primarily driven by low-quality or obscure sources, but rather by inconsistencies in how incident dates are extracted or interpreted.

    \section{Inconsistencies in RIR Naming}
    \label{app:rir-naming}

    Table~\ref{tab:rir-canon-examples} presents concrete RIR name strings alongside the canonical organization we use for aggregation. The examples span (i) brand/legal variants (Amazon), (ii) DBA/store-code boilerplate (Papa John’s Pizza), (iii) product/infra labels (Google Global Cache → Google), (iv) datacenter housekeeping text (o2switch), and (v) regional legal forms (T-Mobile). They illustrate why rule-based normalization is brittle and motivate our LLM-based canonicalization used in the main pipeline.

    \begin{table}[t]
      \centering
      \small
      \setlength{\tabcolsep}{6pt}

      \begin{tabular}{p{0.66\linewidth} p{0.28\linewidth}}
        \toprule
        \textbf{Raw RIR string} & \textbf{Canonicalized} \\
        \midrule
        FTIP004113133 amazon.co.uk & Amazon \\
        AMAZON COMMERCIAL SERVICES PTY LTD & Amazon \\
        "Amazon.com, Inc." & Amazon \\
        \midrule
        PJC - SAN DBA PAPA JOHNS PIZZA - STORE \# 1337 & Papa Johns Pizza \\
        PJ PIZZA DBA PAPA JOHNS PIZZA - STORE \# 1333 & Papa Johns Pizza \\
        PJ CLEVELAND,CA DBA PAPA JOHNS PIZZA - STORE \#3077 & Papa Johns Pizza \\
        \midrule
        Google UK Limited & Google \\
        Google Global Cache & Google \\
        \midrule
        o2switch Datacenter Ip-Range-5 -- Location Datacenter 2 - O2S-PA02 -- ------ o2switch Datacenters ------ Abuse : abuse@o2switch.fr Support : support@o2switch.fr ------ www.o2switch.net ------ & o2switch \\
        o2switch Datacenter Ip-Range-6 -- Location Datacenter 3 - O2S-PA03 -- ------ o2switch Datacenters ------ Abuse : abuse@o2switch.fr Support : support@o2switch.fr ------ www.o2switch.net ------ & o2switch \\
        \midrule
        T-Mobile Austria GmbH & T-Mobile \\
        T-Mobile Czech Republic a.s. & T-Mobile \\
        \bottomrule
      \end{tabular}
        \caption{Raw RIR strings and their canonicalized organization names.}
        \label{tab:rir-canon-examples}
    \end{table}

    \section{LLM-based Organization Name Canonicalization}
    \label{app:llm-canon}

      \paragraph{Examples of Canonicalization Errors and Borderline Cases}
    \label{app:llm-canon-examples}

    Table~\ref{tab:canon-errors} lists representative examples where the
    LLM-based canonicalization made clear mistakes, as well as cases where it was
    unclear whether the output should be considered strictly wrong or an
    acceptable simplification.

   \begin{table}[t]
    \centering
    \small
    \setlength{\tabcolsep}{12pt}
    \begin{tabular}{p{0.4\linewidth}p{0.4\linewidth}}
    \specialrule{1.2pt}{0pt}{0pt}
    Original RIR string & Canonicalized name \\
    \specialrule{0.7pt}{0pt}{0pt}
    \multicolumn{2}{l}{\textbf{Clear mistakes}} \\
    \specialrule{0.7pt}{0pt}{0pt}
    360 NW 27TH STREET TENANT LLC & 360 NW 27TH STREET \\
    \midrule
    Smart Final & Smart \\
    \midrule
    This Space is statically assigned ITK-Net GmbH Lindenstr.\ 4 24619 Bornhved & This Space \\
    \midrule
    NESTLE FRAME PO\#4500293534 & Nestle Frame \\
    \midrule
    Connected by GTS Poland & Connected by GTS \\
    \specialrule{0.7pt}{0pt}{0pt}
    \multicolumn{2}{l}{\textbf{Borderline / not clearly wrong}} \\
    \specialrule{0.7pt}{0pt}{0pt}
    Marina Mini Storage and Truck Rental & Marina Mini Storage \\
    \midrule
    HARDER CLETA SCHOOL & HARDER CLETA \\
    \midrule
    WEIMAN PRODUCTS & Weiman \\
    \midrule
    JiNan-gelaimeiKTV-liaihua- & Jinanktv \\
    \midrule
    1-1819VGJ\_ENTERPRISE Client This space has been assigned as STATIC & 1 \\
    \midrule
    CBGNGC2025022538297\_ENTERP -RISE CLIENT This space has been assigned as STATIC & CBGNGC Enterprise \\
    \midrule
    IP Moshonkin Ilia Sergeevich & IP Moshonkin Ilia Sergeevich (PERSON) \\
    \midrule
    HURLEY AND COMPANY & HURLEY AND \\
    \bottomrule
    \end{tabular}
    \caption{Examples of LLM organization-name canonicalization errors.}
    \label{tab:canon-errors}
\end{table}

    \subsection{LLM Few-Shot Instruction.}

    {\itshape
    You are a data cleaning expert. Your only job is to clean organization names.

    \noindent Rules.
    \begin{enumerate}
      \item Remove legal suffixes (Inc, LLC, Ltd, GmbH, Corp, etc.).
      \item Remove location indicators unless essential.
      \item Remove internal codes, unit identifiers, and meaningless tokens.
      \item Keep only the business/entity name sufficient to uniquely identify the organization.
    \end{enumerate}

    \noindent Examples:

    \noindent Input: ``Amazon Services Inc.'' \\
    Output: Amazon

    \vspace{0.3em}
    \noindent Input: ``Meta Platforms Ireland Ltd.'' \\
    Output: Meta

    \vspace{0.3em}
    \noindent Input: ``Google Corporate - mx1'' \\
    Output: Google

    \vspace{0.3em}
    \noindent Input: ``LB ALUMINIUM BERHAD SELANGOR'' \\
    Output: LB Aluminium

    \vspace{0.3em}
    \noindent Input: ``Harbin population information research institute'' \\
    Output: Harbin Population Information Research Institute

    \vspace{0.5em}
    \noindent Respond with only the cleaned name. No explanations and no extra text.
    }

\section{CISSM–RIR Mapping Examples and Failure Modes}
\label{app:rir_mapping_eval}

\tabref{tab:mapping-failures} lists mismatches from the CISSM–RIR clustering, highlighting the core challenge of name-based canonicalization at scale. Even clearly distinct institutions can appear deceptively similar in text (e.g., \emph{University of Utah} vs.\ \emph{Utah State University}).
\begin{table}[h]
\centering
\small
\begin{tabular}{p{0.4\linewidth} p{0.4\linewidth} r}
\toprule
\textbf{Organization (intended)} & \textbf{Mistaken Mapping} & \textbf{IP Size} \\
\midrule
Glofox & Globax & 2 \\
Southern Hills Eye Care & Southern Hills Health & 8 \\
Utah Pathology & Utah Neurological & 8 \\
Summit Behavioral Healthcare & Summit Mental Health & 8 \\
Calcium Products & Acid Products & 16 \\
Ferguson Medical Group (FMG) & Ferguson Financial Group & 16 \\
Sourcegraph & Stylegraph & 20 \\
British Broadcasting Corporation Pension Scheme & BT Pensions Scheme & 32 \\
Institute of Metallurgy and Materials Science of the Polish Academy of Sciences & Institute of Physics of the Polish Academy of Sciences & 128 \\
Security Service of Ukraine & State Service of Ukraine for Emergencies & 256 \\
Belarusian Police & Bangladesh Police & 256 \\
Saint Xavier University & Xavier University & 669 \\
Merseyside College & Wirral Metropolitan College & 1032 \\
Webster University & Webster College & 2349 \\
University of Utah & Utah State University & 526128 \\
\bottomrule
\end{tabular}
\caption{Mapping errors observed during CISSM–RIR clustering across organization sizes.}
\label{tab:mapping-failures}
\end{table}

\end{document}